\documentclass[Afour,sageh,times]{sagej}

\usepackage{moreverb}

\usepackage{hyperref}
\usepackage[anythingbreaks]{breakurl}
\usepackage{graphicx}
\usepackage{pdfpages}
\usepackage{caption}
\usepackage{xspace}
\usepackage{listings}
\usepackage{xfrac}
\usepackage{color}
\usepackage{mathptmx}

\newcommand\BibTeX{{\rmfamily B\kern-.05em \textsc{i\kern-.025em b}\kern-.08em
T\kern-.1667em\lower.7ex\hbox{E}\kern-.125emX}}

\def\volumeyear{2020}

\newcommand{\bq}{\begin{equation}}
\newcommand{\eq}{\end{equation}}
\newcommand{\bytes}{\mbox{bytes}\xspace}
\newcommand{\byte}{\mbox{byte}\xspace}
\newcommand{\second}{\mbox{s}\xspace}

\newcommand{\flop}{\mbox{flop}\xspace}

\newcommand{\cycle}{\mbox{cy}\xspace}
\newcommand{\cycles}{\mbox{cy}\xspace}

\newcommand{\bit}{\mbox{bit}\xspace}

\newcommand{\GBS}{\,\mbox{G\byte/\second}\xspace}

\newcommand{\GFS}{\mbox{G\flop/\second}\xspace}
\newcommand{\TFS}{\mbox{T\flop/\second}\xspace}

\newcommand{\GHZ}{\mbox{GHz}\xspace}

\newcommand{\FB}{\mbox{\flop/\byte}\xspace}

\newcommand{\eos}{~.}
\newcommand{\cma}{~,}

\newcommand{\AM}{\mbox{$\mathbf{A}$}\xspace}
\newcommand{\BM}{\mbox{$\mathbf{B}$}\xspace}
\newcommand{\CM}{\mbox{$\mathbf{C}$}\xspace}

\begin{document}

\runninghead{Ernst et al.}

\title{Performance Engineering for Real and Complex Tall \& Skinny Matrix Multiplication Kernels on GPUs}

\author{Dominik Ernst\affilnum{1}, Georg Hager\affilnum{1}, Jonas Thies\affilnum{2}, and Gerhard Wellein\affilnum{1}}

\affiliation{\affilnum{1}Erlangen Regional Computing Center (RRZE), 91058 Erlangen, Germany\\
\affilnum{2}German Aerospace Center (DLR), Simulation and Software Technology}

\corrauth{Dominik Ernst, Erlangen Regional Computing Center, Martensstra{\ss}e 1, 91058 Erlangen, Germany}

\email{dominik.ernst@fau.de}

\begin{abstract}
 General matrix-matrix multiplications (GEMM) in ven\-dor-supplied BLAS
  libraries are best optimized for square matrices but often
  show bad performance for tall \& skinny matrices, which are much
  taller than wide. NVIDIA's current CUBLAS implementation delivers
  only a fraction of the potential performance (as given by the roof\/line model)
  in this case. We describe the challenges and key properties
  of an implementation that can achieve perfect performance. We
  further evaluate different approaches of parallelization and thread
  distribution, and devise a flexible, configurable mapping scheme.  A
  code generation approach enables a simultaneously flexible and
  specialized implementation with autotuning.  This results in perfect
  performance for a large range of matrix sizes in the domain of
  interest, and at least 2/3 of maximum performance for the rest
  on an NVIDIA Volta GPGPU.
\end{abstract}


\maketitle

\section{Introduction}

\subsection{Tall \& Skinny Matrix Multiplications}

The general matrix-matrix multiplication (GEMM) is an essential
linear algebra operation used in many numerical algorithms and
hardware vendors usually supply an implementation that is perfectly
optimized for their hardware. In case of NVIDIA, this is part of
CUBLAS (\cite{cublas}). However, since these implementations are
focused on mostly square matrices, they often perform poorly for
matrices with unusual shapes.

This paper covers two types of matrix multiplications with \emph{tall
\& skinny} matrices, i.e.,  matrices that are much taller than they
are wide. We define \emph{skinny} as having in the range of $[1, 64]$
columns, and \emph{tall} as having more than $10^6$ rows. Both
types of multiplications involve the two tall \& skinny matrices $\AM$ and
$\BM$, with sizes $K \times M$ and $K \times N$, respectively,
and $K$ being the long dimension. The small dimensions $M$ and $N$ form a
small matrix $\CM$ with size $M \times N$.

\begin{figure}
\centering
\includegraphics[scale=0.5]{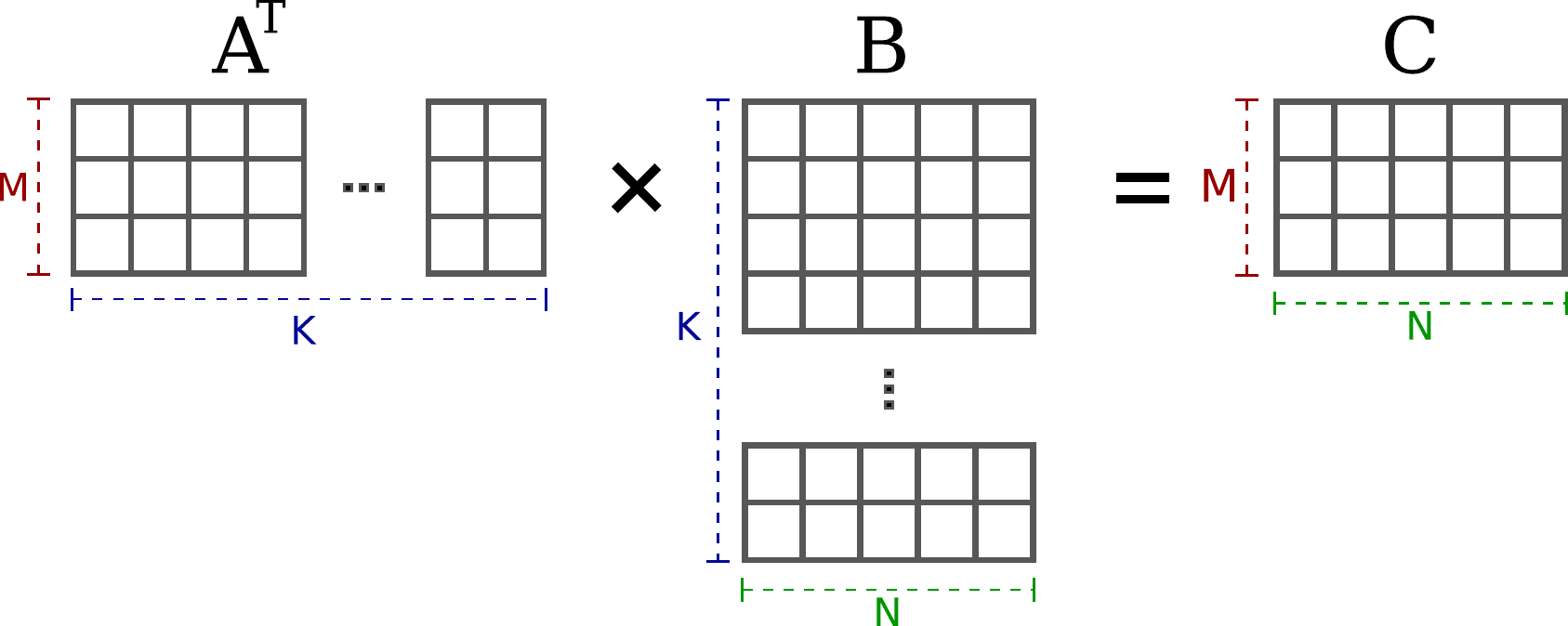}
\caption{The \emph{TSMTTSM} operation $\mathbf{A}^T\,\mathbf{B} =  \mathbf{C} $ with $\mathbf{A}$ and $\mathbf{B}$ being tall \& skinny matrices. Note that $\mathbf{A}$ is transposed in the illustration.} \label{fig:tsmttsm}
\end{figure}

\begin{figure}
\centering
\includegraphics[scale=0.5]{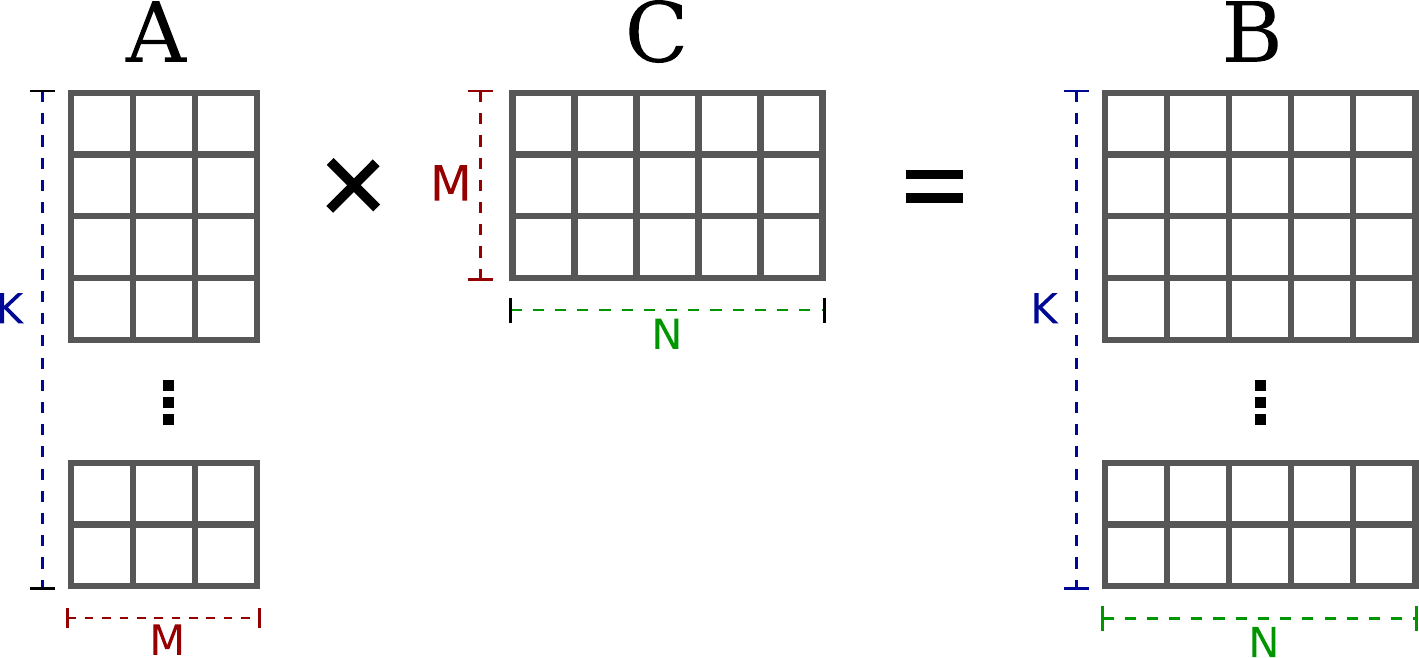}
\caption{The \emph{TSMM} operation $\mathbf{A}^T\,\mathbf{B} =  \mathbf{C} $ with $\mathbf{A}$ and $\mathbf{B}$ being tall \& skinny matrices.} \label{fig:tsmm}
\end{figure}

The two variants are shown in Figures~\ref{fig:tsmttsm} and \ref{fig:tsmm}:
The \emph{Tall \& Skinny Matrix Transposed times Tall \& Skinny Matrix} (\emph{TSMTTSM})
multiplication $\mathbf{A}^T\,\mathbf{B} = \mathbf{C}$ and the
\emph{Tall \& Skinny Matrix times Matrix} (\emph{TSMM}) multiplication
${\mathbf{A} \mathbf{C} = \mathbf{B}}$\@. 

We are interested in a highly efficient implementation of these
operations using double precision real and complex data types on the
NVIDIA Volta GPGPU, used nowadays in many HPC systems.

\subsection{Application}

Row-major tall \& skinny matrices are the result of combining several
vectors to block vectors. \emph{Block Vector Algorithms} are linear
algebra algorithms that compute on multiple vectors simultaneously for
improved performance.  For instance, by combining multiple,
consecutive \emph{sparse matrix-vector} (SpMV) multiplications to a
\emph{sparse matrix-multiple-vector} (SpMMV) multiplication, the
matrix entries are loaded only once and used for the multiple vectors,
which reduces the overall memory traffic and consequently
increases performance of this memory-bound operation. This has first
been analytically shown in~\cite{Gropp99} and is used in many
applications; see, e.g.,~\cite{blockjada,Kreutzer:2018}.

The simultaneous computation on multiple vectors can also be used to
gain numerical advantages. This has been shown for block vector
versions of the Lanzcos algorithm (see~\cite{blocklanzcos}), of the
biconjugate gradient algorithm (see~\cite{blockcg}), and of the
Jacobi-Davidson Method (see~\cite{blockjada}), each of which use block
vectors to compute multiple eigenvectors simultaneously. Many such
algorithms require multiplications of block vectors. For example, both the
\emph{TSMTTSM} (\(\AM^T\BM\)) and \emph{TSMM} (\(\AM\CM\)) occur in
classical Gram-Schmidt orthogonalization of a number of vectors
represented by $\BM$ against an orthogonal basis $\AM$.

\subsection{Roof{}line Model}

We use the roof{}line model by~\cite{roofline} to obtain an upper
limit for the performance of these kernels. In all cases, each of the
three matrices has to be transferred between the memory and the chip
at least once. Even though the directions of data transfers
differ between the kernels, the total data volume does not, as GPUs
generally do not need a write-allocate transfer.  Therefore the
computational intensity is the same for both kernels if $M$ and $N$
are the same. $2MNK$ floating point operations are performed in a
matrix-matrix multiplication, so for double precision the arithmetic
intensity assuming $K \gg M,N$ and $M = N$ is
\begin{equation}\label{eq:ic}
  \begin{alignedat}{2}
  I_D \quad &  \quad \; =  & \frac{2MNK}{(MK + NK + MN) \times 8} & \frac{\flop}{\byte} \\
  &\stackrel{K \gg M,N}{\approx} & \frac{2MN}{(M + N)\times 8} &\frac{\flop}{\byte} \\
  &\;\; \stackrel{M=N}{=}\;\; & \frac{M}{8} & \frac{\flop}{\byte}\eos\\
  \end{alignedat}
\end{equation}
In this symmetric case, the arithmetic intensity grows linearly with
$M$. We will show measurements only for this symmetric case,
although the nonsymmetric case is not fundamentally different, with
the intensity being proportional to the harmonic mean of both
dimensions and consequently dominated by the smaller number.
If the achievable memory bandwidth is $b_s$ (see below), the model
predicts $P_{max} = \min\left( I \times b_s, P_{peak}\right)$ as the
as an absolute upper performance limit. In the case of complex numbers, the
data volume increases by $2 \times$ and the number of floating-point
operations by $4 \times$, resulting in a doubled arithmetic intensity
$I_Z = \frac{M}{4} \flop / \byte$.

With proper loop optimizations in place, the GEMM is usually
considered a classic example for a compute-bound problem with high
arithmetic intensity. However, at $M,N=1$, the arithmetic intensity of
$1/8 \, \FB$, is far to the left of the roof{}line knee of modern
compute devices (typical values ranging from 5\,\FB\ to 17\,\FB) and
strongly memory bound.  This is not surprising given that a matrix
multiplication with $M,N=1$ is the same as a scalar product.  At the
other end of the considered spectrum, at $M,N=64$, the arithmetic
intensity is $8 \, \FB$, which is close to the roofline knee of a V100
GPU (see below).  Therefore the performance character of the operation
changes from extremely memory bound at $M,N=1$ to simultaneously
memory and compute bound at $M,N=64$. An implementation with perfect
performance thus needs to fully utilize the memory bandwidth at all
sizes and additionally reach peak floating point performance for the
large sizes.  The very different performance characteristics make it
hard to write an optimal implementation for both ends of the spectrum,
i.e., different optimizations and specialization is required for both
cases.

\begin{figure}
  \centering
  \includegraphics[scale=0.7]{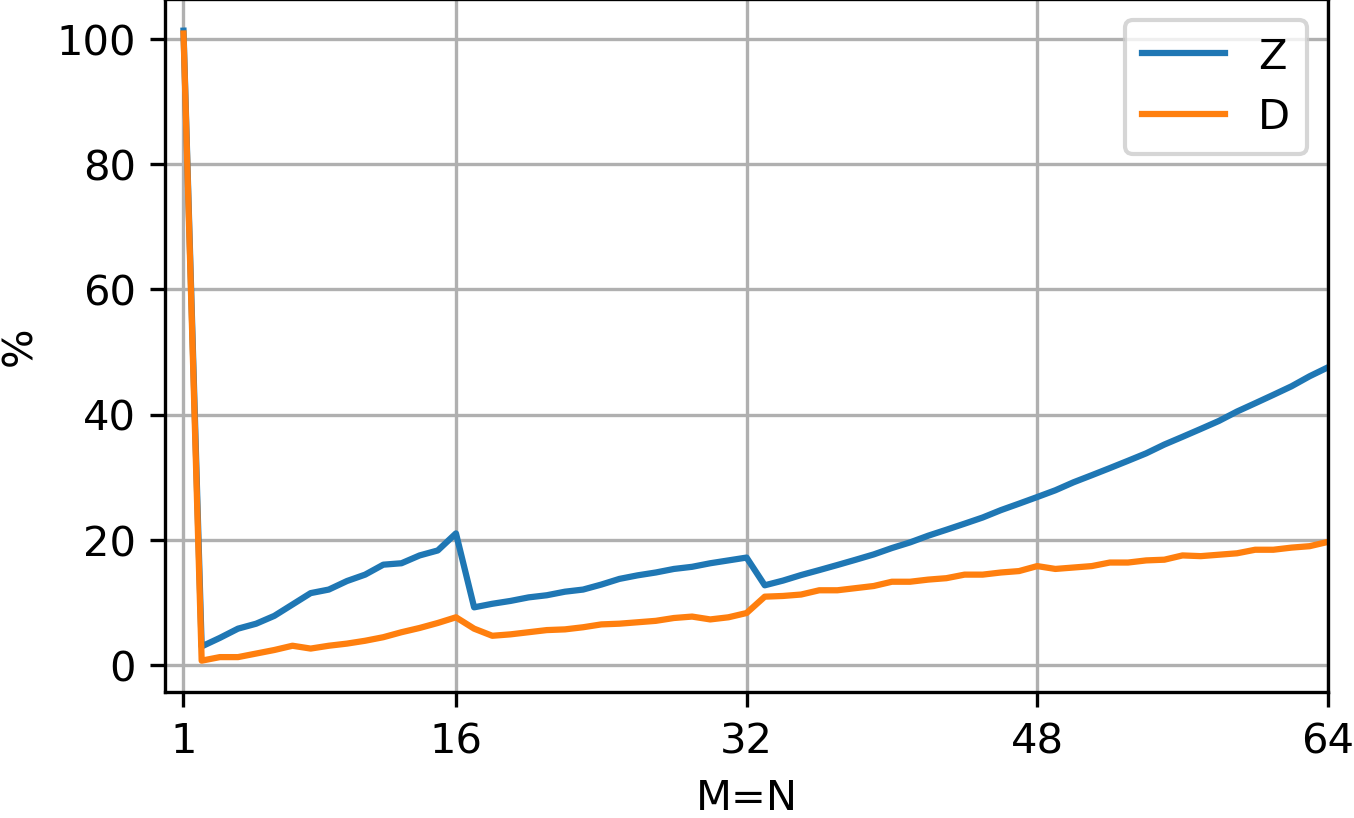}
  \caption{Percentage of roof{}line-predicted performance achieved by
    CUBLAS for the \emph{TSMTTSM} kernel in the range $M=N \in
    [1,64]$, complex (Z) and real (D) double precision, on a Tesla
    V100-PCIe-16GB} \label{fig:cublas_tsmttsm_roofline}
\end{figure}

\begin{figure}
  \centering
  \includegraphics[scale=0.7]{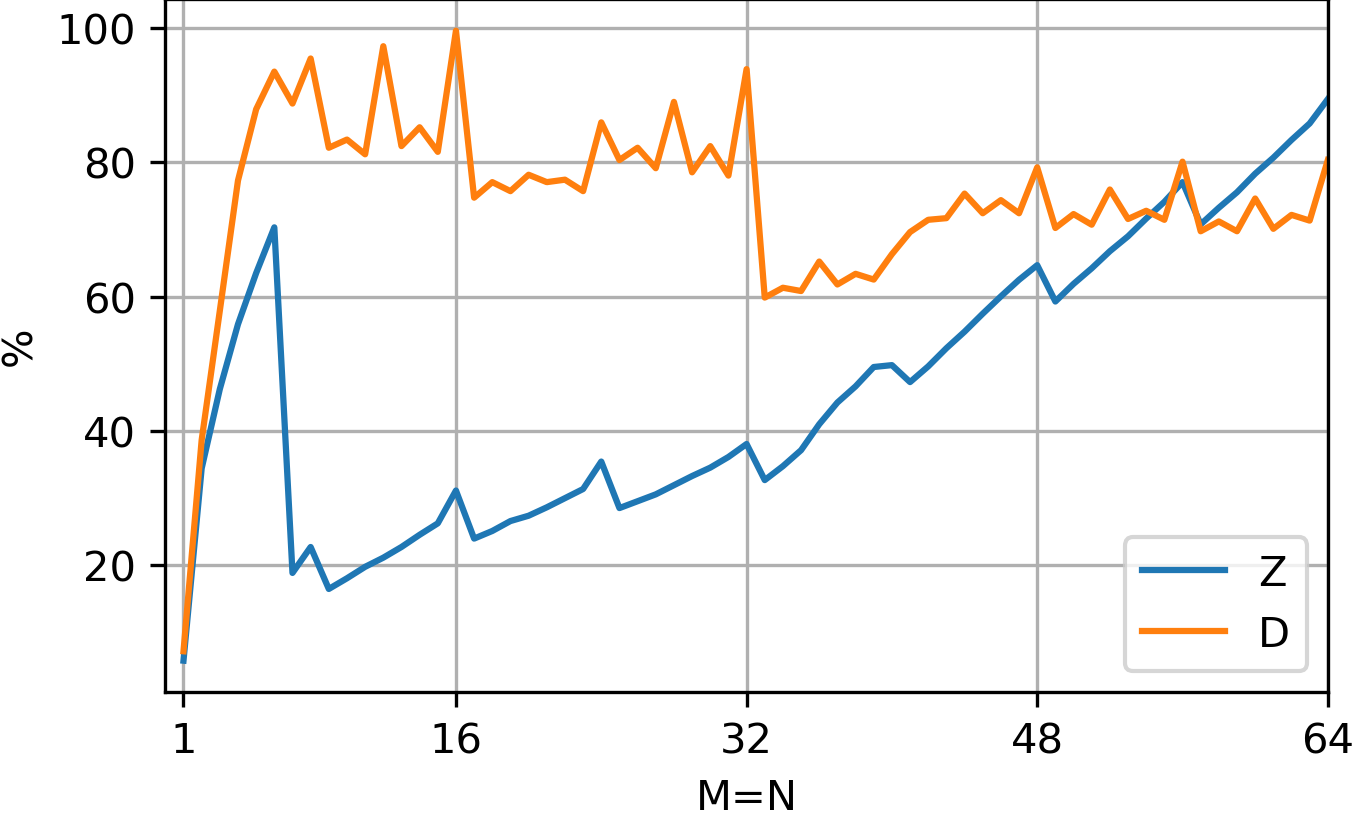}
  \caption{Percentage of roof{}line-predicted performance achieved by
    CUBLAS for the \emph{TSMM} kernel in the range $M=N \in
    [1,64]$, complex (Z) and real (D) double precision, on a Tesla
    V100-PCIe-16GB} \label{fig:cublas_tsmm_roofline}
\end{figure}

It is possible to judge the quality of an implementation's performance
as the percentage of the roof{}line limit. This metric is shown for
CUBLAS in Figures~\ref{fig:cublas_tsmttsm_roofline} and
\ref{fig:cublas_tsmm_roofline}, where the ratio of measured and
roofline performance is plotted as a function of the matrix width.
There is very little performance improvement headroom for CUBLAS'
\emph{TSMM} implementation for real-valued matrices, but there is some
opportunity for complex matrices.  For the \emph{TSMTTSM} kernel,
there is a $2 \times$ to $50 \times$ gap to the upper limit, apart
from $M,N=1$, where NVIDIA obviously implemented a special case
handling. Similarly to the BLAS nomenclature, we use the shorthand
``D'' for double precision real values and ``Z'' for double precision
complex values.

\subsection{Contribution}

This paper presents the necessary implementation techniques to achieve
near-perfect performance for two tall \& skinny matrix-matrix
multiplication variants on an NVIDIA V100 GPGPU with real- and
complex-valued matrices.

To this end, two parallel reduction schemes are implemented and
analyzed as to their suitability for small matrices.

A code generator is implemented that produces code for specific
matrix sizes and tunes many configuration options specifically to that
size.  This allows to exploit regularity where size parameters allow
it, while still generating the least possible overhead where they do
not. As a result, our implementation outperforms state-of-the-art
vendor implementations for most of the parameter range. 

\subsection{Related Work}
This work is an extended version of~\cite{Ernst:2019}\@.  In
comparison to that paper we have added a different variant of
matrix-matrix multiplication (\emph{TSMM}), added a more in depth
performance analysis, extended the analysis to double precision
complex data types, and examined a new \emph{TSMTTSM} thread mapping
scheme.

\emph{CUBLAS} is NVIDIA's BLAS
implementation. The GEMM function interface in BLAS only accepts
column-major matrices. Treating the matrices as transposed column
major matrices and executing $\AM \BM^T$ for the \emph{TSMTTSM} operation and
$\CM \AM$ for \emph{TSMM} are equivalent operations.

\emph{CUTLASS} (\cite{cutlass}) is a collection of primitives for
multiplications especially of small matrices, which can be composed in
different ways to form products of larger matrices. One of these is
the \verb|splitK| kernel, which additionally parallelizes the inner
summation of the matrix multiplication for increase parallelism for the
\emph{TSMTTSM} kernel. We adapted the ``\verb|06_splitK_gemm|''
code sample from the library for benchmarking.

\subsection{Hardware}

In this work we use NVIDIA's V100-PCIe-16GB GPGPU (Volta architecture)
with CUDA 10.0. The hardware data was collected with our own
CUDA micro benchmarks, which are available at~\cite{cuda-benches} together with
more detailed data.

\emph{Memory Bandwidth.}
Whereas the \emph{TSMM} operation has a read and a write stream and
fits well to the ``scale'' kernel from the STREAM benchmarks
(\cite{McCalpin1995}), the \emph{TSMTTSM} is read-only. We thus use a
thread-local sum reduction to estimate the achievable memory bandwidth
$b_s$ (see Table~\ref{tab:membw}). Read-only has a much higher maximum
ceiling of about 880\,\GBS, compared to 820\,\GBS for a ``scale''
kernel. Maximum bandwidth is only attainable with sufficient parallelism, either
through high occupancy or instruction level parallelism (ILP) in the
form of multiple read streams, achieved here through unrolling.

\begin{table}
  \small\sf\centering
  \caption{Measured memory bandwidth on a Tesla V100-PCIe-16GB of a
    read-only kernel with different amount of load parallelism (ILP)
    and occupancies} \label{tab:membw}
  \begin{tabular}{r@{\hskip 1em} | c@{\hskip 0.5em} r@{\hskip 1.0em} r@{\hskip 1.0em} r}
    \toprule
    &  & \multicolumn{3}{c}{ILP, \GBS} \\
    occupancy  &  & 1   & 4    & 16   \\
    \midrule
    1 block, 4 warps &  & 3.0 & 10.1 & 16.3 \\
    6.25\%           &  & 228 & 629  & 815  \\
    12.5\%           &  & 419 & 824  & 877  \\
    25\%             &  & 681 & 872  & 884  \\
    50\%             &  & 834 & 884  & 887  \\
    100\%            &  & 879 & 891  & 877  \\
\bottomrule
\end{tabular}
\end{table}

\emph{Floating-Point Throughput.}
The V100 can execute one 32-wide double precision (DP) floating point
multiply add (FMA) per cycle on each of its 80 streaming
multiprocessors (SMs) and runs at a clock speed of 1.38\,\GHZ
for a DP peak of $80 \times 32 \times 2 \times
1.38\,\GFS = 7066\,\GFS$.
One SM quadrant can process one
instruction that is 32 warp lanes wide every four cycles at a latency of eight
cycles. Full throughput can already be achieved with a single warp per
quadrant if instructions are independent.

\emph{L1 Cache.}
The L1 cache plays is instrumental in achieving the theoretically
possible arithmetic intensity. Though load and DP FMA instructions
have the same throughput of $1/\cycle/SM$, the actual L1 cache
bandwidth of one 128-\byte cache line per cycle means that the actual
load instruction throughput is dependent on the number of touched
cache lines.  For example, a 32-wide, unit-stride DP load touches 2
cache lines and therefore takes two cycles. For that access pattern,
the floating point to load instruction ratio would need to be at least
2:1 to attain peak performance.

\emph{Shared Memory.}
The Shared Memory uses the same physical structure on the chip as the L1 cache.
It has the same bandwidth, but lower access latency than the L1 cache.

\section{General Implementation Strategies}

\subsection{Code Generation}

A single implementation cannot be suitable for all matrix sizes. In
order to engineer the best code for each size, some form of meta
programming is required.  C++ templates allow some degree of meta
programming but are limited in their expressiveness or require
convoluted constructs. Usually the compiler unrolls and eliminates
short loops with known iteration count in order to reduce loop
overhead, combine address calculations, avoid indexed loads from
arrays for the thread-local results, deduplicate and batch loads, and
much more. Direct generation of the intended code offers more control,
however. For example, when using a thread count per row that is not a
divisor of the matrix width, some threads would need to compute fewer
results than others. Guarding \verb_if_ statements have to be added
around these computations that could exceed the matrix size. These can
be omitted wherever it is safe, i.e. all threads compute a valid
value, in order to not compromise performance for even, dividing
thread mappings.  We therefore use a code generating script in python,
which allows to prototype new techniques much quicker and with more
control. Many different parameters can be configured easily and
benchmarked automatically, for example whether leap frogging and
unrolling (see below) are used, how the reduction is performed, and
what thread mapping to set.  The same reasoning for code generation is
made by \cite{codegen}, where it is used to generate small matrix
multiplication kernels for CPUs.

\subsection{Thread Mapping Options}

\begin{figure}
  \centering
  \begin{lstlisting}
  for m = 0...M:
    for n = 0...N:
      for k = 0...K:
        C[m][n] += A[k][m] * B [k][n]
  \end{lstlisting}
  \captionof{lstlisting}{Naive matrix-matrix multiplication (MMM)
    pseudo code for $\CM=\AM^T\BM$.} \label{lst:MMM}
\end{figure}

The parallelization scheme, i.e., the way in which work is mapped to
GPU threads, plays an important role for data flow in the memory
hierarchy. The canonical formulation of an MMM is the three-level loop
nest shown in Listing~\ref{lst:MMM}.

\begin{figure}
  \centering
  \includegraphics{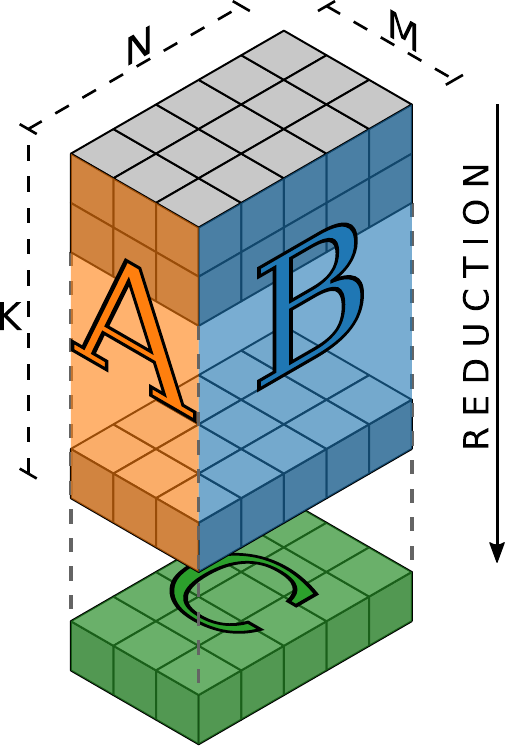}
  \caption{Illustration of the iteration space of the \emph{TSMTTSM}
    operation $\CM=\AM^T \BM$} \label{fig:tsmttsm_cube}
\end{figure}

\begin{figure}
  \centering
  \includegraphics[scale=0.5]{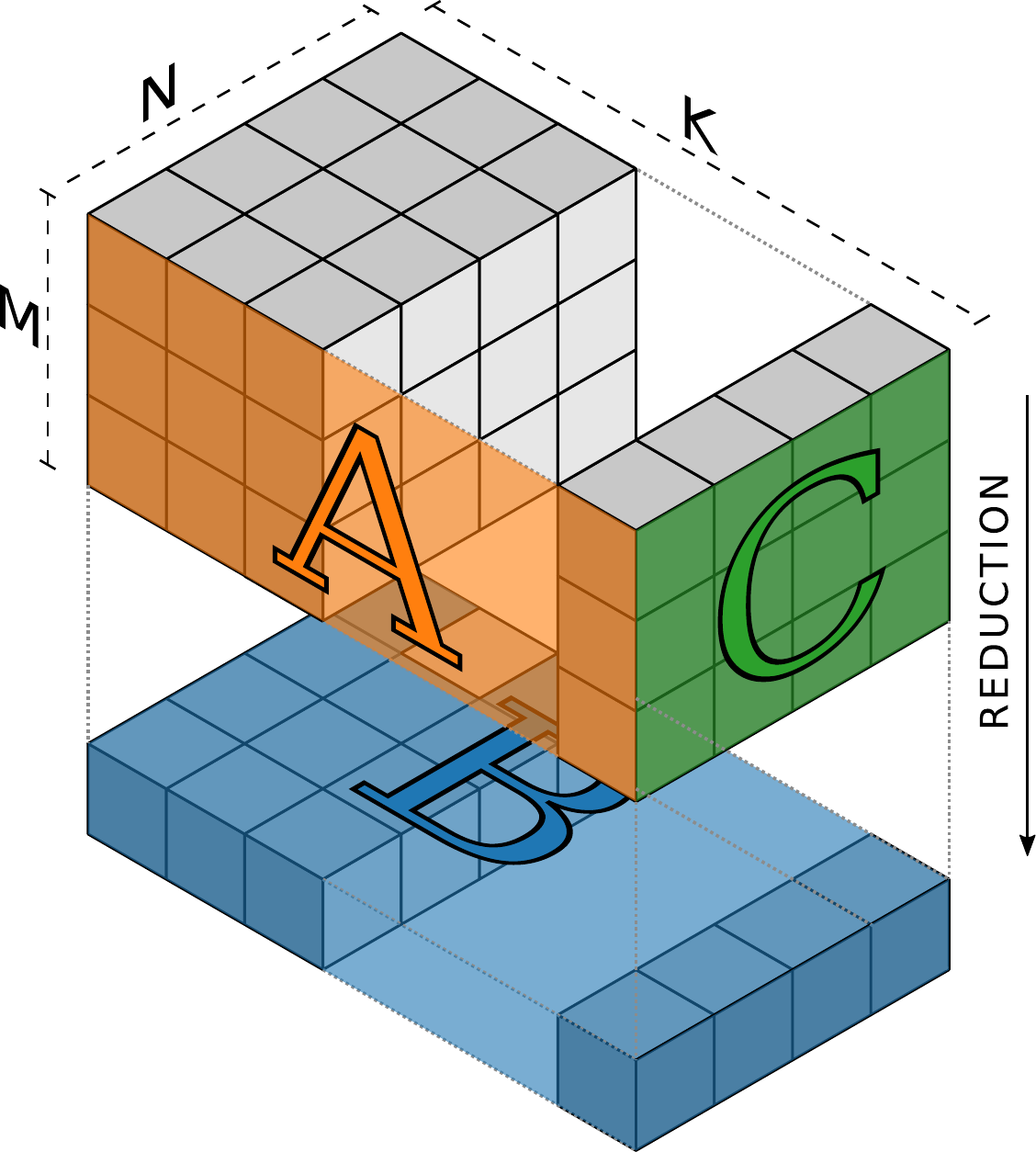}
  \caption{Illustration of the iteration space of the \emph{TSMM}
    operation $\BM=\AM\CM$} \label{fig:tsmm_cube}
\end{figure}

The iteration space of an MMM can be visualized as a cuboid spanned by
the outer product of the two matrices being multiplied. For the
\emph{TSMTTSM} (Figure~\ref{fig:tsmttsm_cube}), the matrices \AM and \BM
span the cube, and reduction along the long axis $K$ results in the
matrix \CM. For the \emph{TSMM} (Figure~\ref{fig:tsmm_cube}), the cube is
flipped on its side, so the the matrices \AM and \CM span the cube and
a reduction along the short side $M$ results in \BM.

This representation allows to visualize the locality of data
transfers. Looking at a slice of the cube perpendicular to
the long $K$ axis spanned by one row of $\AM$ and $\BM$,
as depicted in Figures~\ref{fig:genv1}--\ref{fig:genv8},
shows all the data uses and computations. Each such slice contains $M\times N$ cells, which
correspond to one FMA each, and requires the transfer of one row each of
$\AM$ and $\BM$, causing a data transfer of $M + N$ elements. The
arithmetic intensity associated with the computations in one slice
is the same as for the whole MMM kernel. 
We assume 
perfect caching, i.e., that $\mathbf{A}$ and $\mathbf{B}$ are
transferred from memory just once and reused as many times as necessary
throughout the calculation.

The fastest way to reuse values is
to use a register and have the thread the register belongs to
perform all required operations on this data. Data used by multiple
threads can (preferably) be shared in the L1 cache for threads in the
same thread block or in the L2 cache otherwise. This works only
if some spatial and temporal access locality is in place. Therefore,
the mapping of cells, i.e., work, to threads determines which thread
requires what data for its computations and the locality of data
access.

\section{TSMTTSM}

For the \emph{TSMTTSM}, the two outer loops, which are completely independent
and therefore well parallelizable, are usually the target of an
implementation focused on square matrices. For skinny matrices, these
loops are much too short to yield enough parallelism for a GPU. In
consequence, the loop over the long $K$ dimension has to be parallelized
as well, which also involves parallelizing the sum inside the
loop. There are many more terms in the parallel reduction than
threads, so that each thread can first serially compute a thread local
partial sum, which is afterwards reduced to a total sum, like in
Listing~\ref{lst:tsmttsm1}. Here, a so called grid stride loop,
described at \cite{gridstrideloop}, is used to map rows to threads.

\begin{figure}
  \centering
  \begin{lstlisting}
c_local[:][:] = 0

for (k = threadId; k < K; k += gridStride)
  for m = 0...M:
    for n = 0...N:
      c_local[m][n] += A[k][m] * B [k][n]

for m = 0...M:
  for n = 0...N:
    global_reduction(c_local[m][n])
  \end{lstlisting}
  \captionof{lstlisting}{\emph{TSMTTSM} pseudo code, with the $K$ loop
    parallelized as a grid stripe loop. } \label{lst:tsmttsm1}
\end{figure}

\begin{figure}
  \centering
  \begin{lstlisting}
for (k = threadId; k < K; k += gridStride)
  c0_0 += A[k][0] * B[k][0]
  c0_1 += A[k][0] * B[k][1]
  c1_0 += A[k][1] * B[k][0]
  c1_1 += A[k][1] * B[k][1]
  \end{lstlisting}
  \captionof{lstlisting}{\emph{TSMTTSM} pseudo code with parallelized
    $K$ loop, after unrolling the two inner loops (here shown
    exemplarily for $M=N=2$) and mapping array entries to
    variables. The global reduction is omitted for
    brevity.} \label{lst:tsmttsm2}
\end{figure}

\begin{figure}
  \centering
  \includegraphics[scale=0.9]{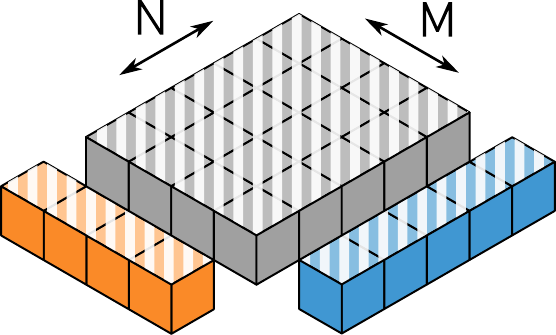}
  \caption{\emph{TSMTTSM}: Parallelization over $K$ only} \label{fig:genv1}
\end{figure}

\begin{figure}
  \centering
  \includegraphics[scale=0.9]{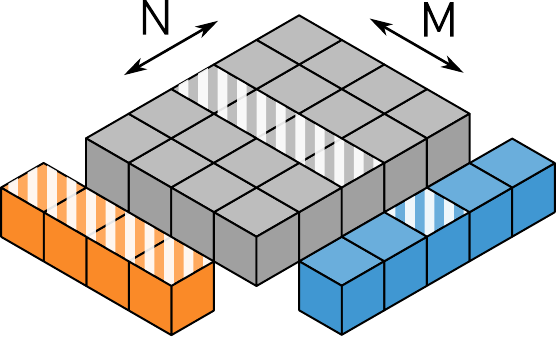}
  \caption{\emph{TSMTTSM}: Parallelization over the $K$ and $N$ loop} \label{fig:genv3}
\end{figure}

\begin{figure} \centering
\includegraphics[scale=0.9]{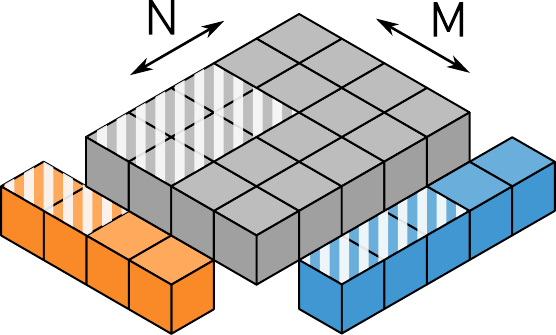}
  \caption{\emph{TSMTTSM}: Parallelization over $K$ and tiling of the two inner loops,
here with tile size $2\times3$} \label{fig:genv8}
\end{figure}

\begin{figure} \centering
\includegraphics[scale=0.9]{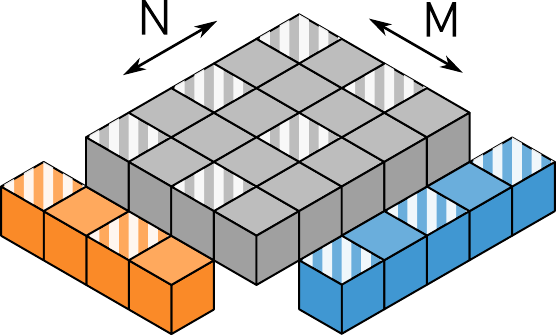}
  \caption{\emph{TSMTTSM}: Parallelization over $K$ and transposed tiling of the two
inner loops, here with tile size
$2\times3$} \label{fig:genvtransposed}
\end{figure}

For data locality, the two small loops have to be moved into the $K$
loop. Since they are short loops with constant loop trip count, they can
be unrolled completely, which also allows instead of indexing into a
local array, to map the intermediates to local variables, like in
Listing~\ref{lst:tsmttsm2}.  Depending on whether and how the two
small loops are parallelized, each thread computes only some of these
$MN$ intermediates. Figures~\ref{fig:genv1} to \ref{fig:genvtransposed}
visualize this by showing a slice of the multiplication cube and which
values a single thread would compute. The number of loads that each
thread has to do are the affected values in the row of $\AM$ and $\BM$,
also visible in the illustrations, while each highlighted cell in the slice
stands for one line in the loop body of Listing~\ref{lst:tsmttsm2}, which
corresponds to one FMA operations and one intermediate variable.

Since the L1 cache is not able to deliver one operand per FMA instruction,
a high FMA-to-load ratio is desirable. This can be achieved by maximizing the
area and the ``squareness'' of the area that is computed by a single
thread. At the same time, more intermediate results per thread
increase the register count, which can limit the occupancy and eventually
lead to spilling.

The approach of only parallelizing the $K$ loop (shown in
Listing~\ref{lst:tsmttsm1} and Figure~\ref{fig:genv1})
easily achieves this goal.  While it maximizes the arithmetic
intensity already in the L1 cache, the $MN$ intermediate results
occupy $2MN$ registers, so the maximum of 256 registers per thread is
already exceeded at $M,N>11$, causing spilling and poor performance.

\begin{figure}
  \centering
  \begin{lstlisting}
for (k = threadId / N; k < K;
     k += gridStride / N)
  n = threadId % N
  for m = 0...M:
    c[m][n] += A[k][m] * B [k][n]
  \end{lstlisting}
  \captionof{lstlisting}{\emph{TSMTTSM} pseudo code, with the $K$ and
    $N$ loop parallelized. The global reduction is
    omitted.} \label{lst:tsmttsm3}
\end{figure}

Parallelizing one of the inner loops as well
(Listing~\ref{lst:tsmttsm3}) leads to the pattern shown in
Figure~\ref{fig:genv3}. The amount of registers required is only $M$
here, so there is no spilling even at $M,N=64$. However, the narrow
shape results in an FMA/load ratio below 1 (i.e., a low arithmetic
intensity in the L1 cache), as values from $\AM$ are used just once
per load.

\begin{figure}
  \centering
  \begin{lstlisting}
midx = (threadIdx / N) % M
nidx = threadIdx % N
for (...)
  for tm = 0...TM:
    for tn = 0...TN:
      m = midx * TM + tm
      n = nidx * TN + tn
      c[tm][tn] += A[k][m] * B [k][n]
  \end{lstlisting}
  \captionof{lstlisting}{\emph{TSMTTSM} pseudo code, with tiled $M$
    and $N$ loop using tile sizes $T_M$ and $T_N$. The global reduction
    and row calculation in the $K$ loop is
    omitted.} \label{lst:tsmttsm4}
\end{figure}

A better approach, which combines manageable register requirements with
a more square form of the tile is to subdivide the two smaller loops into tiles
(see Listing~\ref{lst:tsmttsm4} and Figure~\ref{fig:genv8}). This
mapping also allows for much more flexibility, as the tile sizes can
be chosen small enough to avoid spilling or reach a certain occupancy
goal but also large enough to create a high FMA/load ratio. Tile
sizes that are not divisors of the small loop dimensions can be covered
by generating guarding statements for tile entries that could
possibly overlap to only be executed by threads with a tile index that
does not extend beyond the border of the slice. This is helpful for matrix
dimensions that have few divisors, e.g., prime numbers.

Mapping a continuous range of values to a thread leads to strided loads,
which can be detrimental to performance.  The same entry in two
consecutive threads' partitions is always as far apart as the tile
side length. A more advantageous, continuous load pattern can be
achieved by transposing the threads' tiles, as shown in Figure
\ref{fig:genvtransposed}. The corresponding values in a tile
are now consecutive.

\subsection{Leap Frogging}

On NVIDIA's GPU architectures, load operations can overlap with each other. The
execution will only stall at an instruction that requires an operand from an
outstanding load. The compiler maximizes this overlap by moving all loads to the
beginning of the loop body, followed by the floating-point (FP) instructions that
consume the loaded values. Usually at least one or two of the loads come from
memory and thus take longer to complete than other queued loads, so that
execution stalls at the first FP instruction.  A way to circumvent
this stall is to load the inputs one loop iteration ahead into a separate set of
\emph{next} registers, while the computations still happen on the \emph{current}
values. At the end of the loop, the \emph{next} values become the \emph{current}
values of the next loop iteration by assignment. These assignments are the first
instructions that depend on the loads and thus the computations can happen while
the loads are still in flight.

\subsection{Global Reduction}\label{sec:globred}

After each thread has serially computed its partial, thread-local
result, a global reduction is required, which is considered
overhead. Its runtime depends only on the thread count, though,
whereas the time spent in the serial summation grows linearly with the
row count and therefore becomes marginal for large row counts.
However, as shown by~\cite{jonas_toms}, the performance
at small row counts can still be relevant, as the available GPU memory
may be shared by more data structures than just the two tall \& skinny
matrices, limiting the data set size.

Starting with the \emph{Pascal} architecture, atomic add operations for double
precision values are available for global memory, making global
reductions more efficient than on older systems.  Each thread can just
use an \verb|atomicAdd| of its partial value to update the final results. The
throughput of global \verb|atomicAdd| operations is limited by the
amount of contention, which grows for smaller matrix sizes.  We
improve on this global atomic reduction variant with a local atomic
variant that reduces the amount of global \verb|atomicAdd| operations
by first computing thread-block-local partial results using shared
memory atomics. This is followed by a global reduction of the local
results. Additionally, we opportunistically reduce the amount of
launched threads for small row counts.

\section{TSMM}

\subsection{Thread Mapping}
\begin{figure}
  \centering
  \includegraphics[scale=0.5]{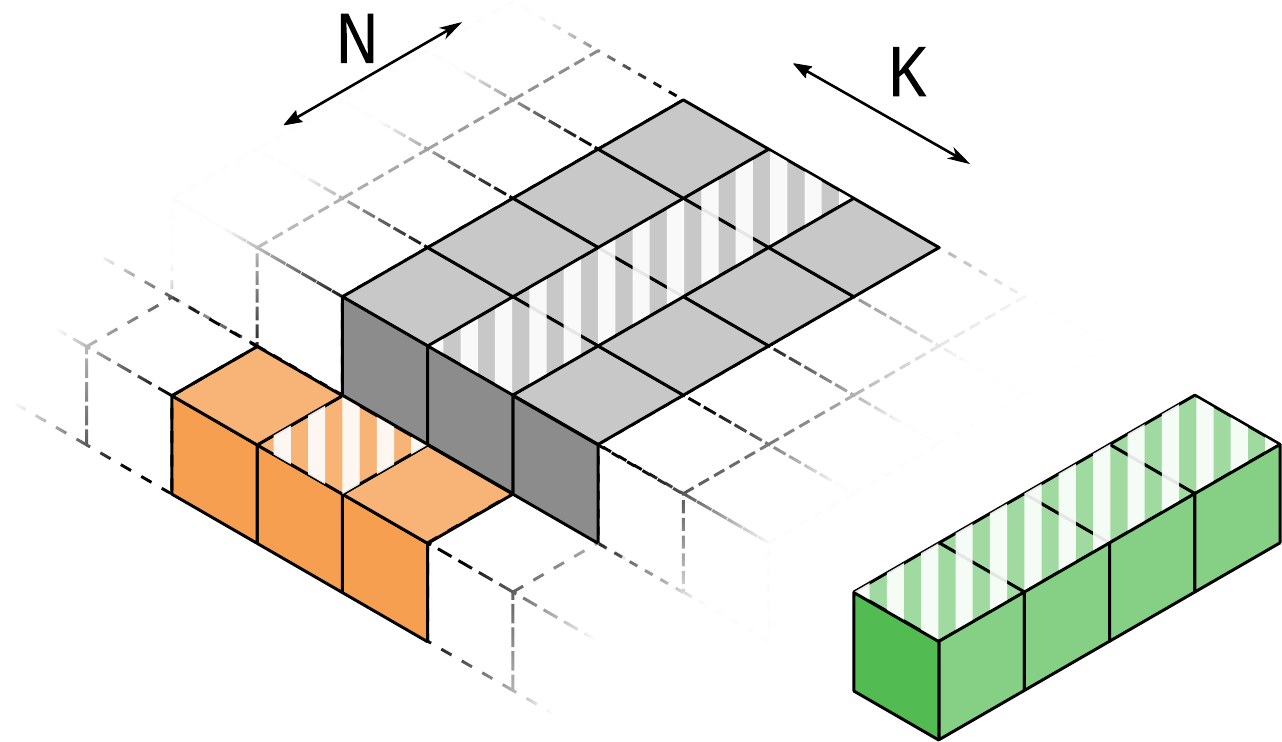}
  \caption{\emph{TSMM}: Parallelization over $K$, a single thread
    computes a full result row of $\BM$. Slice perpendicular to the
    $M$ axis, the (long) $K$ axis extends ``indefinitely'' on both
    sides.} \label{fig:tsmm_mapping_1T}
\end{figure}

\begin{figure}
  \centering
  \includegraphics[scale=0.5]{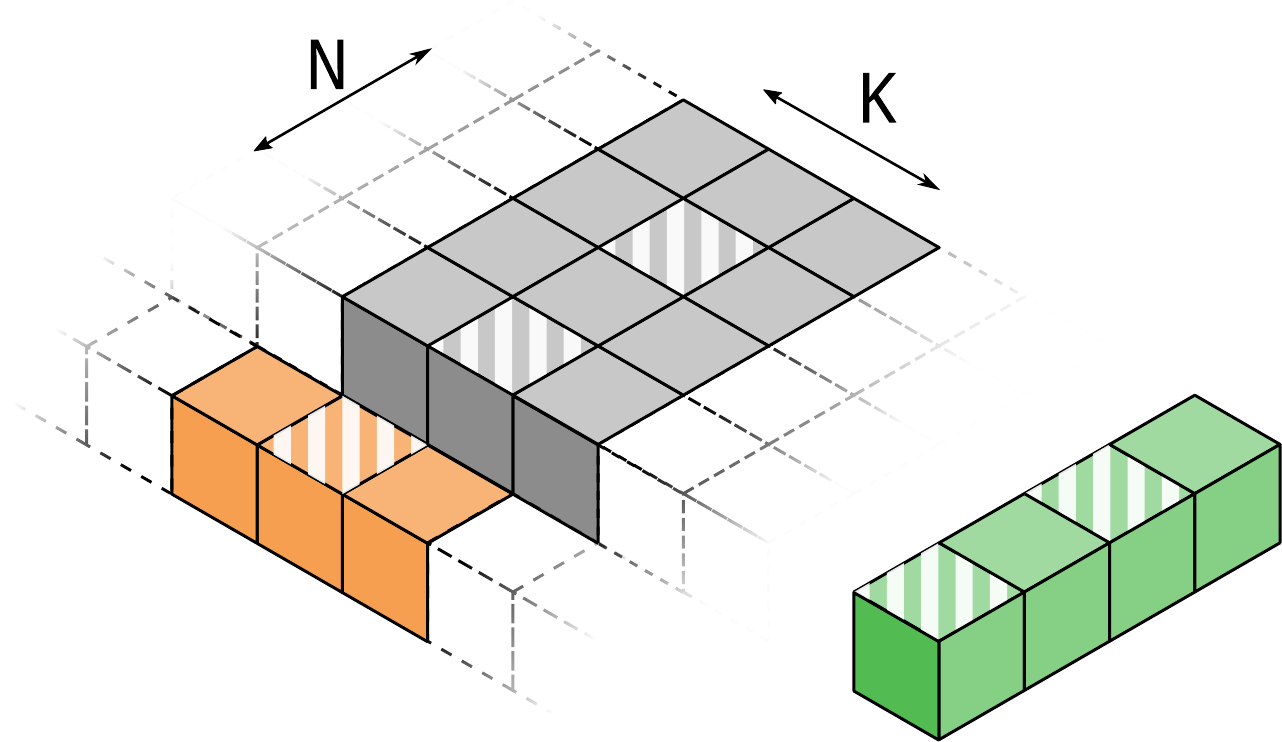}
  \caption{ \emph{TSMM}: Parallelization over $K$ and $N$, two threads
    compute two results each. Slice horizontal to the $M$ axis, the
    $K$ axis extends ``indefinitely'' on both
    sides.} \label{fig:tsmm_mapping}
\end{figure}

\begin{figure}
  \centering
  \includegraphics[scale=0.5]{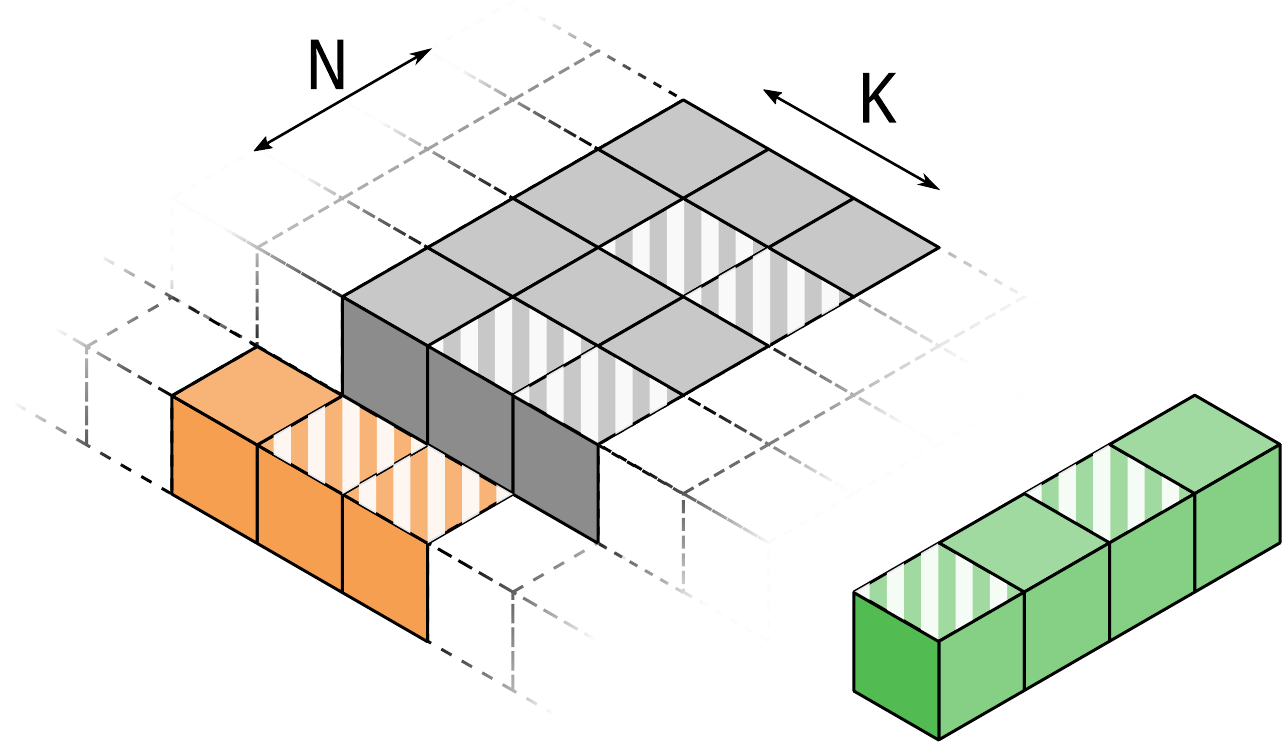}
  \caption{\emph{TSMM}: Parallelization over $K$ and $N$ and $2\times$
    unrolling, two threads compute two results each and on two rows of
    $\BM$ in a single iteration. Slice horizontal to the $M$ axis, the
    $K$ axis extends ``indefinitely'' on both
    sides.} \label{fig:tsmm_mapping_u}
\end{figure}

In contrast to the \emph{TSMTTSM} kernel, the summation is done along the
short $M$ axis, with no need for a global reduction. Though
the short sum could be parallelized, this is not necessary in this
case, as the other two loop dimensions supply sufficient parallelism.
The visualizations in
Figures~\ref{fig:tsmm_mapping_1T}-\ref{fig:tsmm_mapping_u} show
slices perpendicular to the $M$ axis, since this dimension will
not be parallelized.

The first option is to only parallelize over the long $K$ dimension as
shown in Figure~\ref{fig:tsmm_mapping_1T}. Each entry in $\AM$
would be loaded once and then reused out of a register. The $N$ sums
that each thread computes require a $2N$ registers, which
is not a prohibitive number even at $N=64$ but still does
reduce occupancy. A more severe disadvantage are the strided stores.  As
each thread produces and stores a full row of $\BM$, the addresses
stored to by the different threads are far apart, leading to partially
written cache lines.  This in turn causes a write-allocate read
stream of the result matrix $\BM$ to ensure fully consistent cache
lines, thereby reducing the arithmetic intensity of the kernel.

This can be avoided by parallelizing the $N$ loop. Each thread
computes a single result of the output row of $\BM$. Because
consecutive threads compute consecutive results, cache lines are
always written fully and no write-allocate stream is necessary. The
disadvantage is a low compute/load ratio. Each value from $\AM$ is
loaded and used just once in each thread.

A more balanced approach is to have a smaller group of threads compute
on each result row, with a few results computed by each thread. Each
value loaded from $\AM$ is reused multiple times, once for each result
computed by this thread.  Using a transposed mapping as shown in
Figure~\ref{fig:tsmm_mapping}, each thread does not compute
consecutive elements; results computed by threads are
interleaved, so that consecutive elements are written and the
amount of partial writes is reduced. This works best if the thread count is a
multiple of four, which corresponds to the L1 cache line management
granularity of $32\;\bytes$. If $N$ is not a multiple of four, the
writes will necessarily be misaligned, with some cache lines being
cut. Larger thread counts slightly reduce the impact of cut cache
lines.

\subsection{Data from $\CM$}

Our discussion of thread mappings and data transfers
so far has ignored the entries of the matrix $\CM$. These values are the same for
every index of the $K$ loop.
The fastest would be to load all entries of $\CM$ into registers and
reuse them from there, but this strategy would quickly exceed
the number of available registers even at moderate $M$ and $N$.
Since they are accessed
frequently and all threads in a thread block access similar values,
the contents of $\CM$ should continuously stay in the L1 cache, making
reloads of these values a question of L1 cache bandwidth and not
memory latency. Each load from $\CM$ loads between one to three 128-\byte
cache lines, which would then be used for a single FMA. This
is higher than the sustainable ratio of one 128-\byte cache line per
FMA. A solution is to reuse each value loaded from $\CM$
by unrolling the $K$ loop and pulling the unrolled iterations inside
the $M$ loop. Each iteration over $K$ loads the same values of $\CM$,
which can subsequently be used for multiple iterations per load.

The loads from $\CM$ can also be sped up by using the shared memory to
cache these loads. Threads in a thread block collaboratively load the
contents of $\CM$ into the shared memory at the beginning of the
kernel. The loop over $K$ is parallelized with a grid stride loop,
where only as many threads as necessary for full occupancy are
launched. Each kernel instantiation then computes on multiple rows of
$\BM$. Therefore, loading $\CM$ into shared memory can be amortized
over many rows.

On the V100, the shared memory has the same bandwidth as the L1 cache,
given that they occupy the same hardware structure. However, shared memory
accesses guarantee cache hits, as they avoid conflict misses with other data.
They also have a lower latency, since no tags have to be checked.

\section{Results: TSMTTSM}

\subsection{Transposition and Leap Frogging}

\begin{figure}
  \centering
  \includegraphics[scale=0.7]{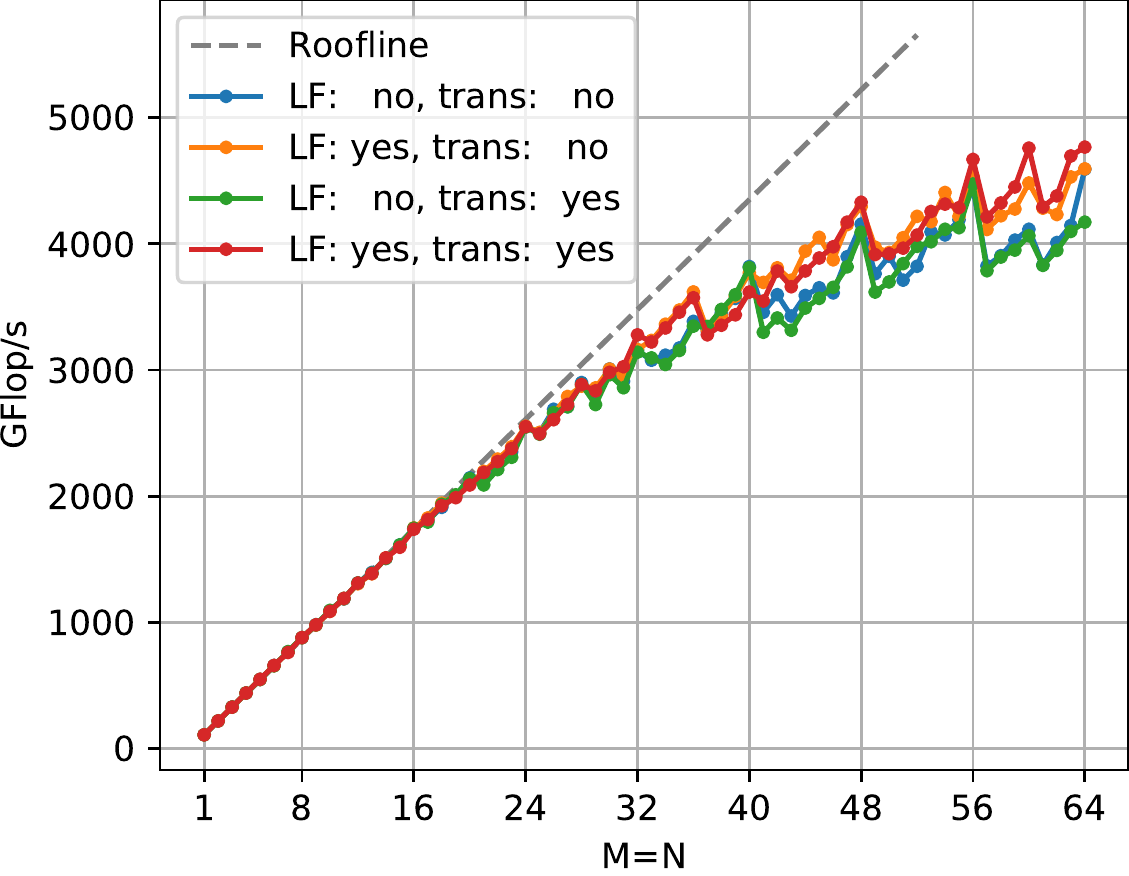}
  \caption{Performance comparison of real-valued double-precision \emph{TSMTTSM}
    vs.\ quadratic tile size
    with $K=2^{29}/M$ on the V100 across the four different permutations of using leap
    frogging (LF) and transposed mapping (trans). The best performance for
    each matrix size and configuration is shown. The arithmetic peak performance
  of the device is $7.066\,\TFS$.} \label{fig:perfperm}
\end{figure}

An exhaustive search was used to find the best tile size and
configuration for each matrix size. The simpler mapping schemes are subsets
of the tiled mapping. E.g., the mapping in Figure~\ref{fig:genv3} corresponds
to a tilesize of $M\times 1$. Figure~\ref{fig:perfperm} shows
the performance of the four configurations of using leap frogging and
a transposed mapping.  The performance agrees with the roof\/line prediction
(dashed line) perfectly until $M,N=20$. Until $M,N=36$, the best
performance stays within 95\% of the limit.  Beyond that, the growing
arithmetic intensity does not translate into a proportional speedup
anymore, although the performance is still about a factor of two away
from peak. The best variants plateau at about
4700\,\GFS, or \sfrac{2}{3} of peak.  Both variants using leap
frogging are clearly faster, but the transposed mapping is only
a bit faster if leap frogging is used. This is in contrast to experiences
with the \emph{Kepler} GPU architecture, where strided loads are slower, and this kind
of transformation is more beneficial. The best tile size changes when
leap frogging is used as it requires more registers.

\subsection{Tile Sizes}

\begin{figure}
\centering
\includegraphics[scale=0.7]{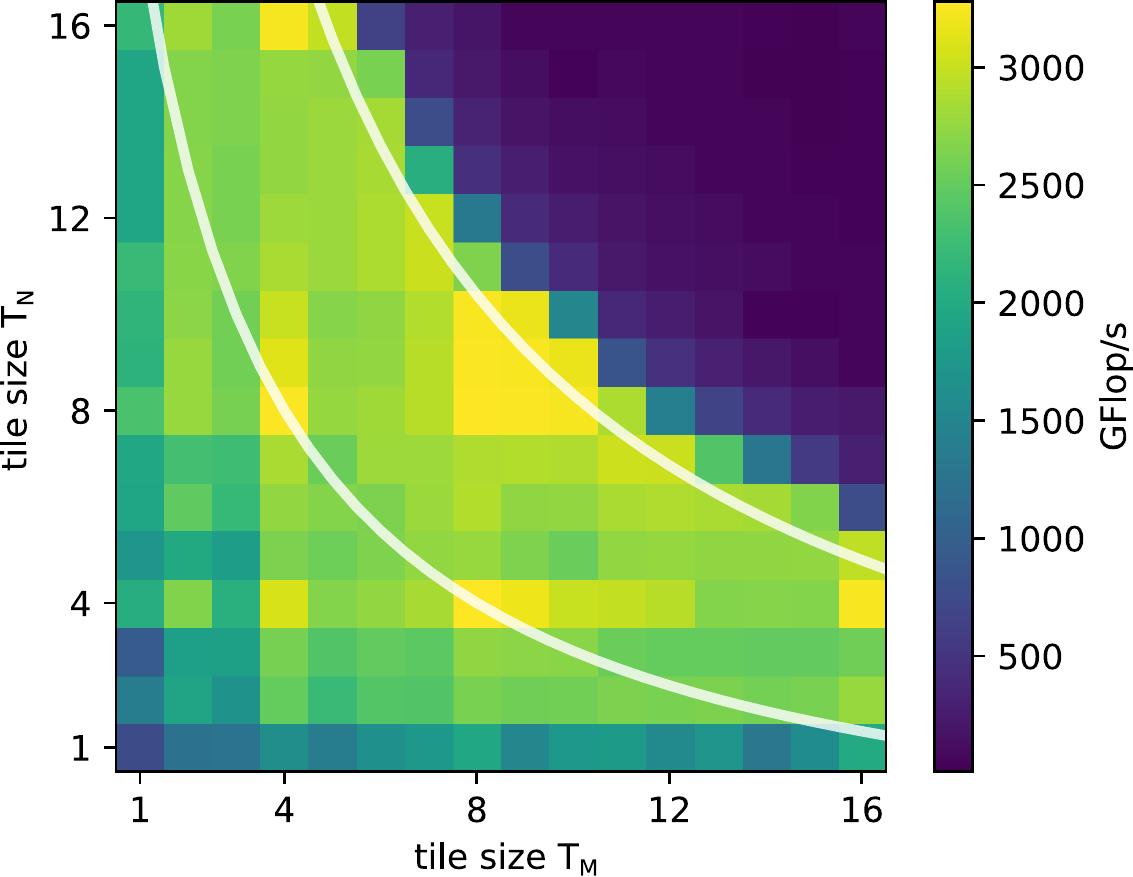}
\caption{Performance of \emph{TSMTTSM} for $M,N=32$ and $K=2^{29}/M$
  vs.\ tile sizes in $M$ and $N$ directions,
  using real-valued double-precision matrices, with leap frogging and
  transposed mapping. The two white lines are defined by
  $2 \times (T_MT_N + 2(T_M+T_N) + 8) = R$, with $R=128, 256$
  to mark approximate
  boundaries of register usage.} \label{fig:tileperf}
\end{figure}

Figure~\ref{fig:tileperf} shows the dependence of performance on the tile
sizes $T_M$ and $T_N$ for the case $M,N=32$ with leap frogging and
transposed mapping. Performance drops off sharply if the tile sizes
become too large and too many registers are used. The number of
registers can be approximated by $2 \times (T_MT_N + 2(T_M+T_N) + 8)$,
which accounts for the thread-local sums ($T_MT_N$), loaded values
$(T_M+T_N)$, and eight registers for other purposes. Leap frogging
introduces a factor of two for the number of loaded values (for 
\emph{current} and \emph{next} values), and double precision values generally
require two 32-\bit registers for an overall factor of two. The graph shows the
iso-lines of 128 and 256 registers, which represent the occupancy drop
from 25\% to 12.5\% at 128 registers and the onset of spilling at 256
registers.

The best-performing tile sizes generally sit on or just below these
lines, maximizing the area of the tile for a given occupancy. The
dimensions are largely symmetric but not perfectly so, as threads are
mapped to tiles in $M$ direction first. There are clear patterns
favoring powers of two as those are divisors of the matrix size 32
and avoid both the overhead of guarding statements and idle threads.

\subsection{Analysis}

According to the roof\/line model, at $M=N=64$
the upper performance limit is
\bq
P=\frac{64}{8}\frac{\flop}{\byte} \times 880\,\GBS = 7060\,\GFS\cma
\eq
which is almost exactly the $P_{Peak}$ of $7066\,\GFS$.
However, our implementation cannot realize the roof\/line-predicted
performance, and instead tops out at $4766\,\GFS\approx\sfrac{2}{3}P_{Peak}$.
The reason for the limitation is memory latency,
which can be shown by a simple model:
Whereas the memory latency for an idle memory interface measured with
a pointer chasing benchmark (see~\cite{cuda-benches}) is only
435\,\cycles, this latency increases as the load on the memory interface
increases. For the values in Table~\ref{tab:membw}, it is possible to
calculate corresponding latency values according to Little's Law via
\bq\label{eq:little}
T_\ell=\frac{fN \times 8\,\byte}{b}\cma
\eq
with $f$ being the clock frequency, $N$
the thread count and $b$ the memory bandwidth. For the unloaded case in the
first row of Table~\ref{tab:membw} (ILP=1), the latency
according to (\ref{eq:little}) is $T_\ell\approx 470\,\cycles$,
which matches the measured pointer chasing latency quite well.
The bandwidth of $b=681\,\GBS$ at 25\% occupancy
in the fourth row roughly corresponds to the highest observed memory
bandwidth, based on the computational intensity,
for $M,N=64$, and result in $T_\ell\approx 664\,\cycles$
of memory latency.

The best tile size without leap frogging is $11\times8$, which
requires $11 \times 8 = 88$~FMA operations. These can be computed on a
single quadrant in $88\times 4\,\cycles=352\,\cycles$. At this large tile
size, the register requirements of at least $2\times11\times8=176$
registers allow to run only eight warps, i.e., two warps per quadrant,
simultaneously on a SM. One warp doing 352\,\cycles of compute
work finishes earlier than the other warp waiting for 664\,\cycles for
data from memory. It will then also wait for the next data to be
loaded, which is a period of time where none of the two warps are
issuing floating point operations, and therefore counts as wasted cycles.

Leap frogging does improve the situation, as even with a single warp
the memory latency and compute times can overlap. However, additional
registers are required to hold the data for the next iteration, which
either necessitates smaller tile sizes or reduces occupancy, both of
which are bad for overlapping. Overall, leap frogging is still beneficial,
though.

\begin{figure}
  \centering
  \includegraphics[scale=0.7]{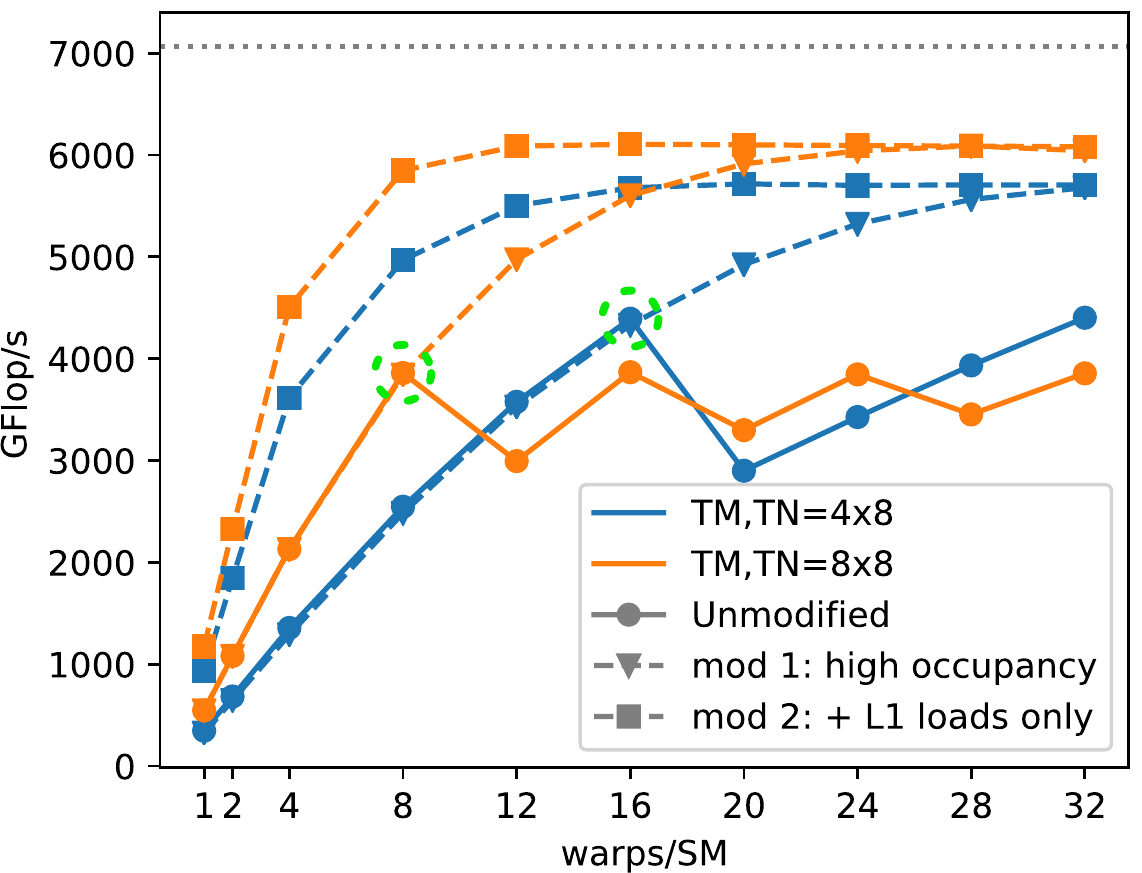}
  \caption{\emph{TSMTTSM} performance vs.\ occupancy of real, correct kernels and two
    modified (incorrect) kernels at tile sizes of $4\times 8$ and $8\times 8$,
    respectively. The first modification reduces
    register count, while the second kernel additionally reduces the data
    set so that it resides in L1 cache. Green circles mark the point
    with the highest performance of the unmodified kernels. (Real-valued
    double-precision matrices)} \label{fig:perfoccupancy}
\end{figure}

Figure~\ref{fig:perfoccupancy} shows an experiment that gives insight
into the relationship of latency and occupancy. A modification of the
generated kernels allows testing the impact of higher occupancies even
for kernels with larger tile sizes, where the high register
requirements usually limit the occupancy to the minimum of eight
warps per SM. Instead of computing $T_M\times T_N$ intermediate results, all
summands are summed up in just two accumulators. This does of course
not compute the correct results any more, but all the instructions and
loaded operands are the same, while reducing the register count so
that 32 warps per SM can run concurrently. Another modification to the
generated kernel introduces a division of the $K$ loop row index by a
large constant. In consequence, all loop iterations compute on data of
very few rows, which makes almost all accesses L1 cache hits with the
corresponding much smaller latency. Repeatedly using the same row is
done in such a contrived way in order to prevent the compiler from
pulling the loads in front of the loop.

With tile sizes of $8\times 4$ and $8 \times 8$, as used in this
experiment, 16 and 8 warps per SM can run concurrently. At these
occupancies (green circles in Figure~\ref{fig:perfoccupancy}),
the respective real kernels (circle symbols) performance is highest,
as the maximum possible number of
thread blocks run concurrently.  With an increased number of launched
thread blocks, the unmodified kernels' performance does not increase
anymore, as additional thread blocks do not run concurrently but are
scheduled in a ``second wave'' of thread blocks. An imbalance in the
number of thread blocks per wave leads to fluctuating performance.

The kernels modified for higher occupancy (triangle symbols) have
the same performance as the unmodified kernels up to these points, but
allow to see the hypothetical speedup if more thread blocks
could run concurrently, which would be possible on a hypothetical
V100 with $4\times$ larger register files.

The performance increase is linear in all cases up to four warps per SM, as this is
the minimum to fill all four quadrants of a SM. For both tile sizes,
the L1 load kernels (square symbols) profit somewhat from a second
warp on each quadrant to overlap the remaining latency and overhead
but quickly saturate at ceilings of $6080\,\GFS$ and $5700\,\GFS$,
respectively, which is not a latency effect any more. The reason for these
lower roofs remains open, but we suspect that it be rooted in limited
instruction throughput.
We noticed that the gap to the device peak performance
matches one missing DP FP operation per four non-DP FP operations,
i.e., integer and load instructions. DP FP operations are supposed to
execute on separate execution units, and so we can only speculate
whether there is a restriction in co-issuing DP FP operations with
integer and load instructions.

The two experiments with the normal, higher latency from memory (triangular
symbols) need many more warps to overlap their longer latency to
eventually saturate at the same level as the L1 load kernels. At least
two to three times larger register files would be required to get
there.  At the same time, it also shows how devastating it would be if
the register files were half as large, a situation that is not dissimilar to
the older \emph{Kepler} GPU architecture, where double the number of execution units
were backed by a similar sized register file.
The larger tile size saturates more quickly,
because it amortizes the same latency over twice the number of 
floating-point operations. Note that in the end, both tile sizes have
a similar real world performance, as the higher possible occupancy of
16 warps per SM compared to 8 warps per SM balances the smaller amount of work
per iteration.

This simple model also helps to explain the rather
small benefits from using the transposed mapping. The transposed mapping
changes the load pattern to contiguous blocks instead of long
strides. This in turn reduces the number of touched cache lines, and
increases the rate at which the L1 cache can serve the outstanding loads
after the data has arrived from memory. However, this rate is only
really a limiter at low FMA/load ratios, or at the beginning of the
floating-point operation phase, where the FP units still
wait for enough registers being filled for uninterrupted operation. The
transposed mapping therefore only gives a small speedup in phase
that is mostly not the limiter, but at the same time also makes
smaller tile sizes more feasible.

On the other hand, the strided access patterns of the nontransposed mapping
touch most cache lines already on the first load, and therefore already
cause most cache misses with the first load. Subsequent loads are cache hits.
With the transposed mapping, with its contiguous blocks of addresses per load,
cache misses are postponed until later loads, which starts the memory latency
penalty later. That is why the configuration
using the transposed mapping without leap frogging performs the worst (see
Figure~\ref{fig:perfperm}). However,
in combination with leap frogging it is faster than the two variants with the
nontransposed mapping.

\subsection{Comparison with Libraries}

\begin{figure}
  \centering
  \includegraphics[scale=0.7]{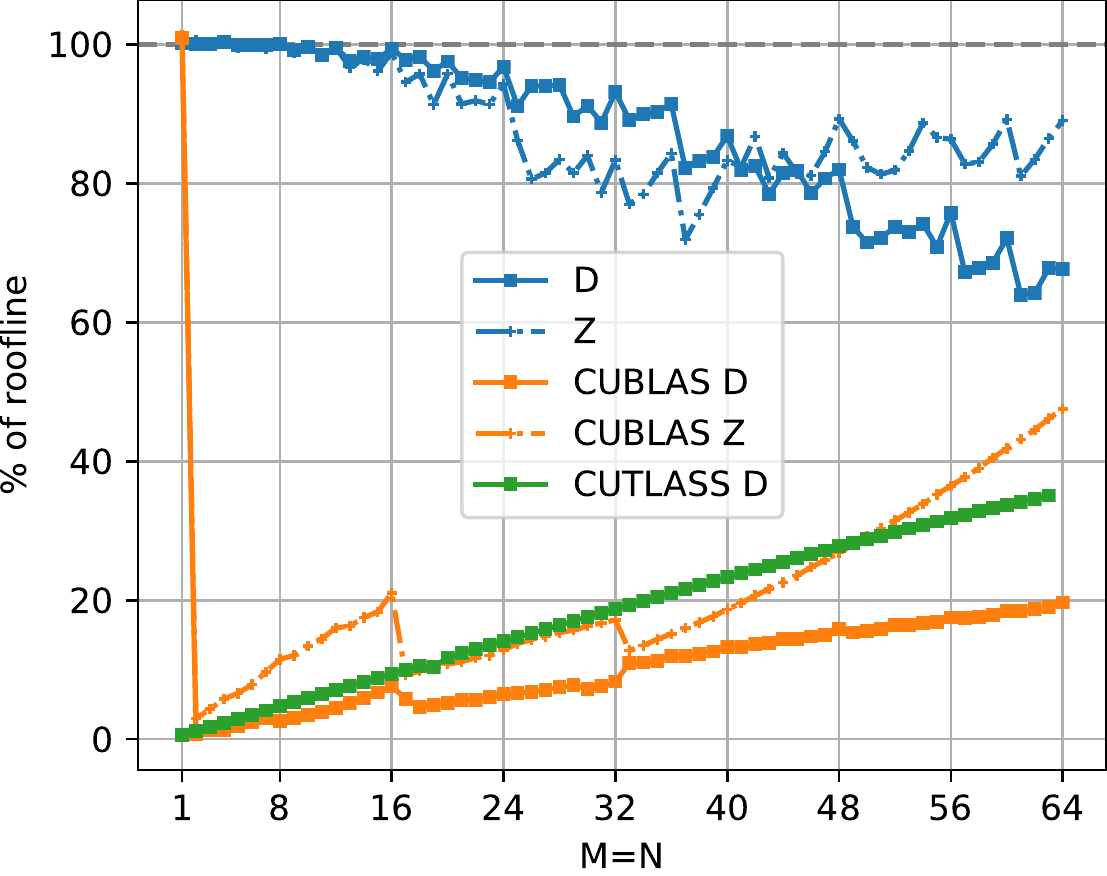}
  \caption{\emph{TSMTTSM} percentage of roof\/line-predicted performance for real (D) and
    complex (Z) double-precision data in comparison with CUBLAS and
    CUTLASS.} \label{fig:bestflop}
\end{figure}

Both CUBLAS' and CUTLASS' performance is far below the potential
performance, except for $M,N=1$, where CUBLAS seems to have a special
detection for the scalar product corner case. The utilization of
potential performance increases as matrices become wider, which makes
them more square and compute bound, bringing them closer to more
standard scenarios.

In contrast, the presented implementation shows full efficiency for
narrow, clearly memory bandwidth limited matrices, and utilization
slightly drops off as matrices become more compute bound. For
complex-valued matrices, the \emph{TSMTTSM} becomes compute bound
already at $M,N=32$. Instruction throughput becomes the limiter much
earlier instead of memory bandwidth and latency, which is why the
utilization drops earlier. With increasing matrix size, it fully
saturates the previously explained lower ceiling due to our speculated
co-issue limitation between double-precision FP instructions and
integer instructions.

\subsection{Impact of Reductions}\label{sec:reductions}

\begin{figure}
  \centering
  \includegraphics[scale=0.6]{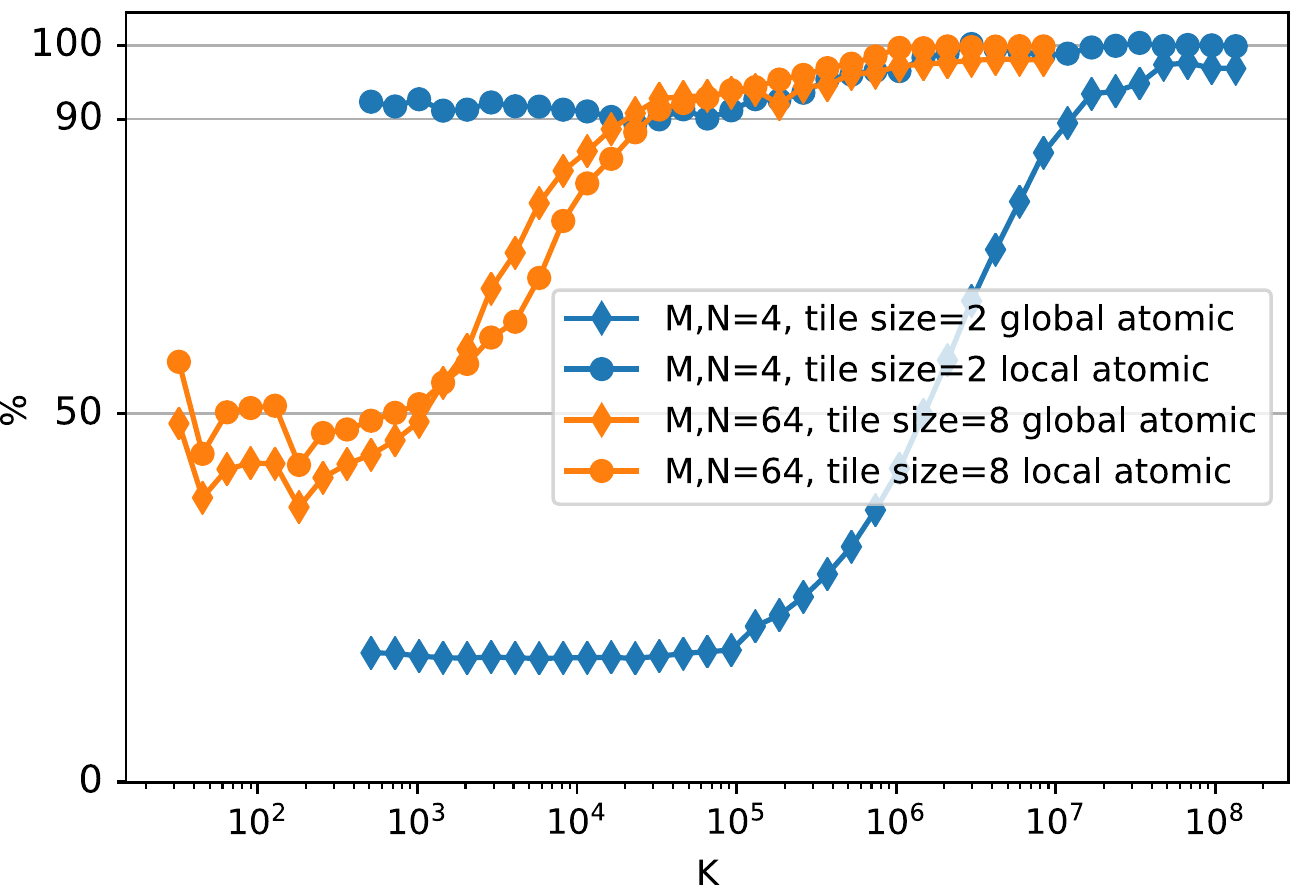}
  \caption{Global reduction impact for \emph{TSMTTSM}:
    Performance when using each
    of the two global reduction variants as the percentage of the
    performance of a kernel without a global reduction, using two different
    matrix widths and tile sizes. (Real-valued double-precision
    matrices).}\label{fig:red_perf}
\end{figure}

Figure~\ref{fig:red_perf} shows the relative performance of our \emph{TSMTTSM}
implementation versus row count with respect to a baseline without any
reduction for a selection of inner matrix sizes and tile sizes,
choosing either of the two reduction methods described in
Section~\ref{sec:globred}.  As expected, the impact of the reduction
generally decreases with increasing row count. The method with only
global atomics is especially slow for the narrower matrices
($M,N=4$). Many threads writing to a small amount of result values
leads to contention and causes a noticeable impact even for a matrix
filling the device memory ($K=10^8$). The local atomic variant drastically
reduces the number of writing threads, resulting in less than 10\%
overhead even for the smallest sizes and near-perfect performance for $K>10^6$.
For the wider matrices, the difference is smaller. The global atomic
version is not as slow because writes spread out over more result
values and the local atomic variant is not as fast because the larger tile size
requires more work in the local reduction. Both variants incur less than $10\%$
overhead just above $K=10^4$, a point where only about $0.2\%$ of the GPU
memory is used.

\section{Results: TSMM}

The described methods and parameters open up a large space of configurations.
Each of the Figures~\ref{fig:tsmmplot1}, \ref{fig:tsmmplot2}, and \ref{fig:tsmmplot3}
shows a cross section of each configuration option by displaying the best-performing
value for each choice for that configuration option.

\subsection{Source of $\CM$}

\begin{figure}
  \centering
  \includegraphics[scale=0.7]{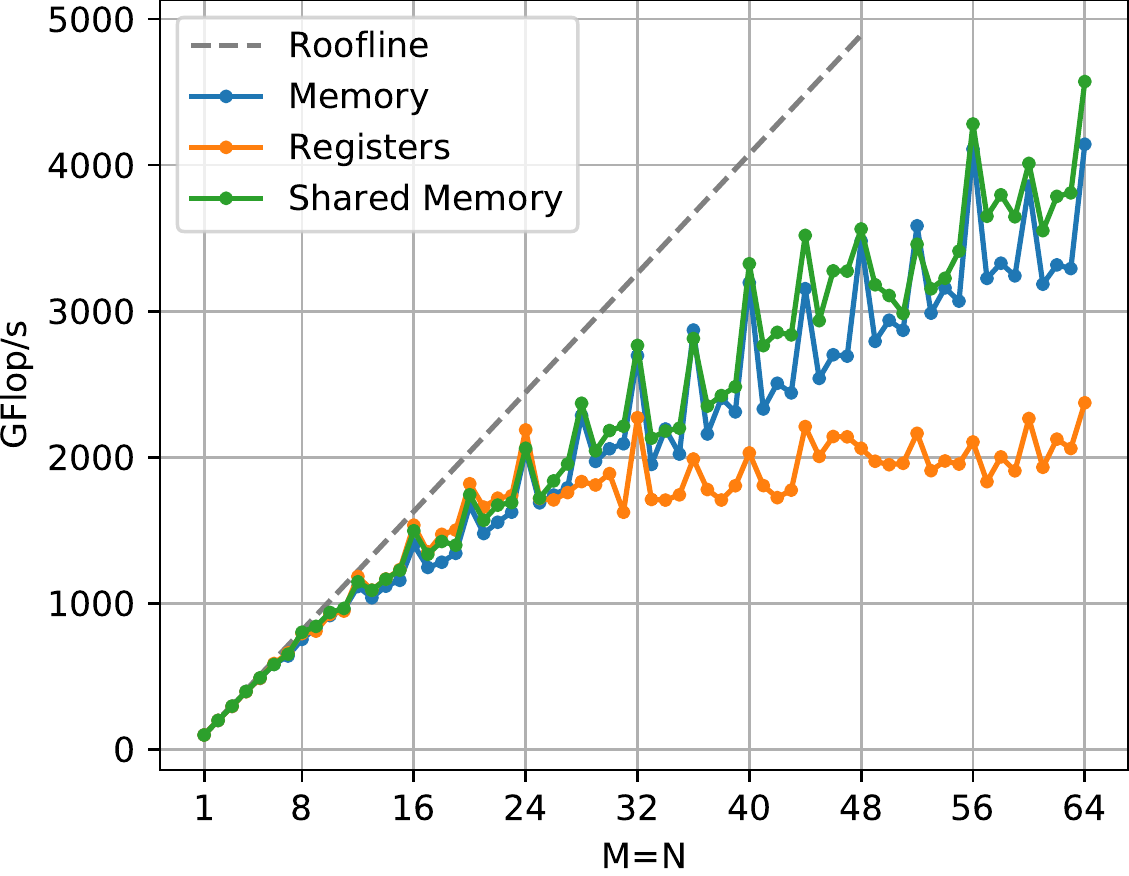}
  \caption{\emph{TSMM} performance comparison at $K=2^{29}/M$ among
    different sources for the matrix $\CM$, showing the
    best-performing configuration of each method and matrix
    width. (Real-valued double-precision matrices).} \label{fig:tsmmplot1}
\end{figure}

The data in Figure~\ref{fig:tsmmplot1} demonstrates that trying to keep the values of
matrix $\CM$ in registers works well only for small $M,N$.
The increasing register pressure at larger sizes reduces occupancy, which is
especially bad if multiple results are computed
per thread.

Reloading values from shared memory consistently has a small
performance advantage especially for sizes that are not multiples of
four, due to a smaller penalty because of misaligned loads. Each
additional cache line that gets touched because of misalignment costs
an additional cycle.

\subsection{Unrolling}

\begin{figure}
  \centering
  \includegraphics[scale=0.7]{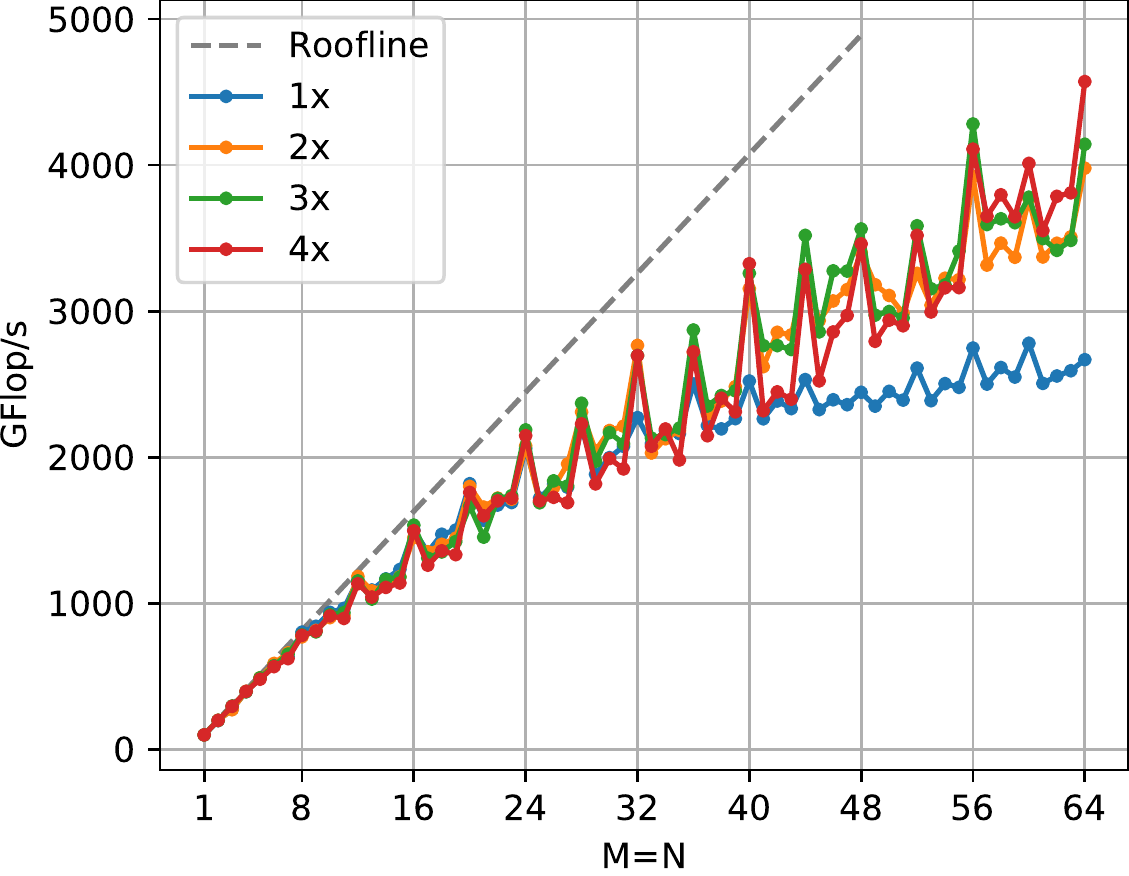}
  \caption{\emph{TSMM} performance comparison of different degrees of
    unrolling at $K=2^{29}/M$, showing the best-performing configuration
    for each unrolling depth ($1,\ldots,4$)
    and matrix width. (Real-valued double-precision
    matrices).} \label{fig:tsmmplot2}
\end{figure}

Although there is little improvement with further unrolling beyond
$2\times$, as Figure~\ref{fig:tsmmplot2} shows, unrolling at least
once shows a clear speedup compared to no unrolling. Without
unrolling, the shared memory bandwidth would limit the performance
due to the high
ratio of shared memory loads to FP DP instructions, and its latency
could not be hidden as well with FP DP instructions from further
iterations.  Generally, a similar reasoning as for the \emph{TSMTTSM}
kernel applies, where computing more results per thread and higher
unrolling counts increase the number of floating-point operations per
iteration but also decrease the occupancy that would be needed to
overlap the memory latency.

\subsection{Thread Count}

\begin{figure}
  \centering
  \includegraphics[scale=0.7]{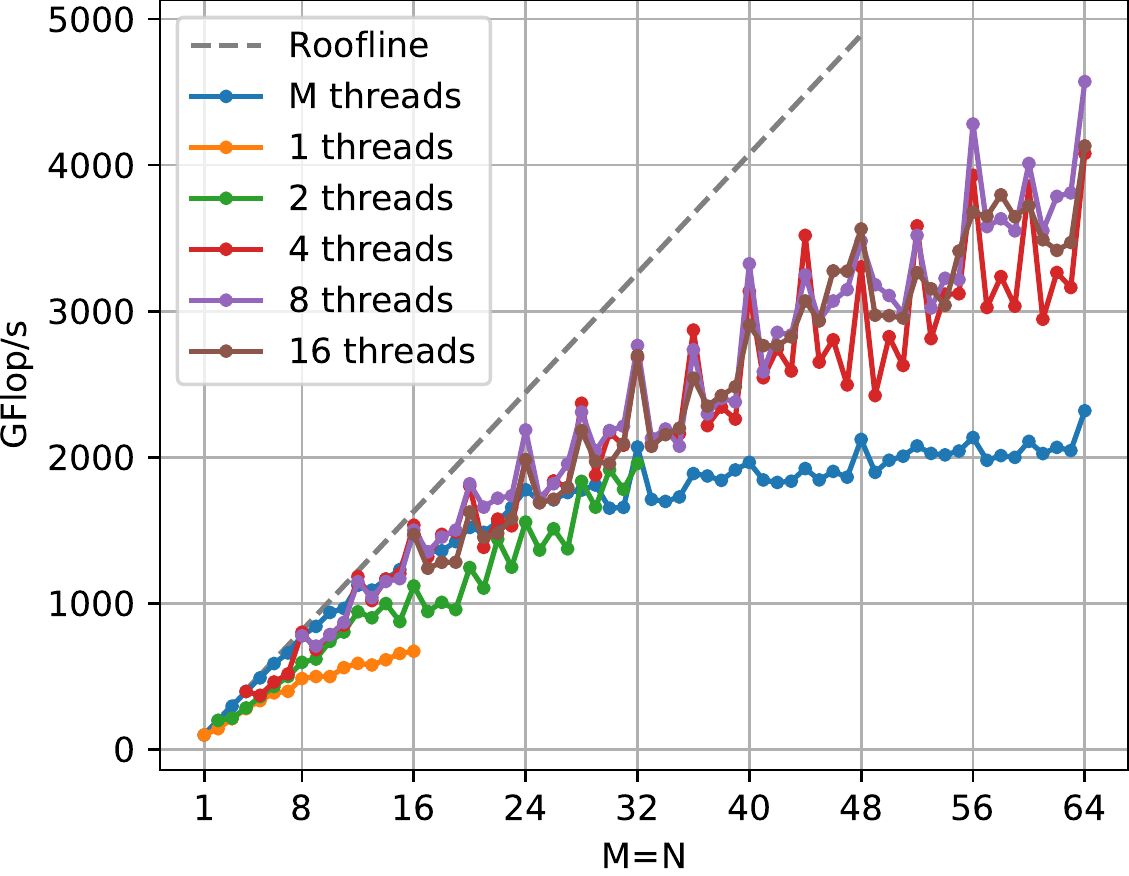}
  \caption{\emph{TSMM} performance comparison of different thread
    counts per row at $K=2^{29}/M$, showing the best-performing configuration for
    each thread count and matrix width. (Real-valued double-precision matrices).}
  \label{fig:tsmmplot3}
\end{figure}

Fewer threads per row mean more work per thread. For large
matrix sizes, this can result in huge kernels with high register
requirements, which is why Figure~\ref{fig:tsmmplot3} does not show
measurements for the whole matrix size range for one and two threads
per row. These two thread counts are the slowest variants, as they
show the effects of strided writes the most. With four threads writing
consecutive values, there is at least a chance of writing a complete
32-\byte cache line sector. The difference between 4, 8 or 16 threads is
not large, although the larger thread counts perform slightly more
consistently (i.e., with less fluctuation across $M$).

The performance analysis for \emph{TSMM} shows a clear
preference for the small matrix dimension $M=N$ to be a multiple of
four. For this case, all writes of computed data to the matrix $\BM$ are
aligned to $4 \times 8 \,\byte = 32 \,\byte$, which is the management
granularity for L1 cache lines and the cache line length for the L2
cache. With this alignment, cache lines are fully written and there is
no overhead for write allocation from memory. Misalignment is the
major performance hurdle for matrix widths that are not multiples of
four.

\subsection{Comparison with Libraries}

\begin{figure}
  \centering
  \includegraphics[scale=0.7]{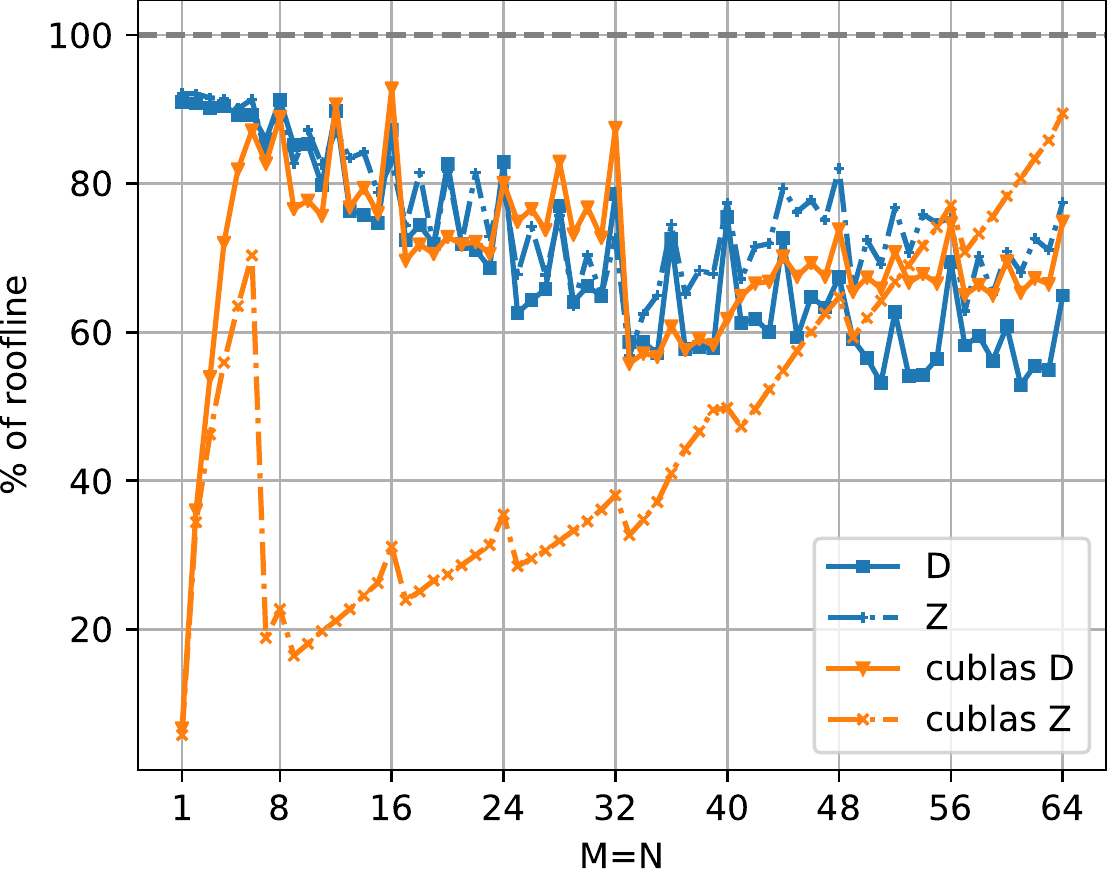}
  \caption{\emph{TSMM} percentage of roof\/line-predicted performance for real
    (D) and complex (Z) double-precision data in comparison with CUBLAS.}
  \label{fig:tsmmplot4}
\end{figure}

Figure~\ref{fig:tsmmplot4} shows that, except for very small $M,N$, 
CUBLAS performs very well for the real-valued \emph{TSMM}
kernel. With increasing width, the development in utilization is
very similar to the presented implementation. Our solution 
works similarly well for complex values matrices, which is not the
case for CUBLAS. Here, a strong performance drop for medium-wide matrices
can be observed.

\section{Conclusion and Outlook}

We have shown how optimize the performance for two
types of multiplication of double-precision, real and complex tall \&
skinny matrices on a V100 GPU. With matrices 
narrower than 32 columns, perfect performance in accordance
with a roof\/line performance model could be achieved.
Over the rest of the skinny range up to a width of 64, 
between 60\% and \sfrac{2}{3} of the potential performance was
attained. 
We used a code generator on top of a range of
suitable thread mapping and tiling patterns, which enabled an
exhaustive parameter space search.  Two different ways to achieve
fast, parallel device-wide reductions for long vectors have been
devised in order to ensure a fast ramp-up of performance already for
shorter matrices.
An in-depth performance analysis was provided
to explain observed deviations from the roof\/line limit.
Our implementation outperforms the vendor-supplied CUBLAS and CUTLASS
libraries by far or is on par with them for most of the observed
parameter range.

In future work, in order to push the limits of the current implementation,
shared memory could be integrated into the mapping scheme to speed up the many
loads, especially scattered ones, that are served by the L1 cache.

The presented performance figures were obtained by parameter search.
An advanced performance model, currently under development, could be fed with
code characteristics such as load addresses and instruction counts generated
with the actual code and then used to eliminate bad candidates much faster.
It will also support a better understanding of performance limiters.

Prior work by us in this area is already part of the sparse matrix toolkit
\emph{GHOST} (\cite{GHOST}) and we plan to integrate the presented work
there as well.\medskip

\begin{funding}
This work was supported by the ESSEX-II project in the DFG Priority Programme SPPEXA.
\end{funding}

\bibliographystyle{SageH}
\bibliography{references}
\end{document}